\documentclass[11pt]{article}

\usepackage{amsmath}
\usepackage{amssymb}
\usepackage{amsthm}
\usepackage{paralist}
\usepackage[misc]{ifsym}
\usepackage{epsfig}
\usepackage{epstopdf}
\usepackage{algorithm}
\usepackage{algorithmic}
\usepackage{subcaption}
\usepackage[export]{adjustbox}
\usepackage[colorlinks=true]{hyperref}
\hypersetup{urlcolor=blue, citecolor=red}
\allowdisplaybreaks

\usepackage[margin=1in]{geometry}

\theoremstyle{definition}

\DeclareMathOperator{\Unif}{Unif}
\DeclareMathOperator{\SNR}{SNR}
\DeclareMathOperator*{\argmax}{arg\,max}
\DeclareMathOperator{\SO}{SO}

\newcommand{\rev}[1]{{{#1}}}

\makeatletter
\newcommand{\@LN}[2]{}
\makeatother

\title{Expectation-Maximization for Structure Determination Directly from Cryo-EM Micrographs}

\author{
Shay Kreymer\thanks{Corresponding author: \href{mailto:shaykreymer@mail.tau.ac.il}{shaykreymer@mail.tau.ac.il}. School of Electrical and Computer Engineering, Tel Aviv University, Tel Aviv 69978, Israel.}
\and
Amit Singer\thanks{Department of Mathematics and Program in Applied and Computational Mathematics, Princeton University, Princeton, NJ 08544, USA.}
\and
Tamir Bendory\thanks{School of Electrical and Computer Engineering, Tel Aviv University, Tel Aviv 69978, Israel.}
}

\date{}

\begin{document}
\maketitle

%%%%%%%%%%%%%%%%%%%%%%%%%%%%%%%%%%%%%%%%%%%%%%%%%%%%%%%
%             5. ABSTRACT
%%%%%%%%%%%%%%%%%%%%%%%%%%%%%%%%%%%%%%%%%%%%%%%%%%%%%%%

\begin{abstract}
A single-particle cryo-electron microscopy (cryo-EM) measurement, called a micrograph, consists of multiple two-dimensional tomographic projections of a three-dimensional (3-D) molecular structure at unknown locations, taken under unknown viewing directions. All existing cryo-EM algorithmic pipelines first locate and extract the projection images, and then reconstruct the structure from the extracted images. However, if the molecular structure is small, the signal-to-noise ratio (SNR) of the data is very low, making it challenging to accurately detect projection images within the micrograph. Consequently, all standard techniques fail in low-SNR regimes. To recover molecular structures from measurements of low SNR, and in particular small molecular structures, we devise an approximate expectation-maximization algorithm to estimate the 3-D structure directly from the micrograph, bypassing the need to locate the projection images. We corroborate our computational scheme with numerical experiments and present successful structure recoveries from simulated noisy measurements.
\end{abstract}

%%%%%%%%%%%%%%%%%%%%%%%%%%%%%%%%%%%%%%%%%%%%%%%%%%%%%%
%                   6. BODY
%%%%%%%%%%%%%%%%%%%%%%%%%%%%%%%%%%%%%%%%%%%%%%%%%%%%%%

% Only the first word and proper nouns of section titles should be capitalized.
% The title of section 1:
\section{Introduction}

Cryo-EM is an increasingly popular technology in structural biology for elucidating the 3-D structure of biomolecules~\cite{nogales2016development, bai2015cryo}. In a cryo-EM experiment, individual copies of the target biomolecule are dispersed in a thin layer of vitreous ice. Then, a 2-D tomographic projection image, called a \textit{micrograph}, is produced by an electron microscope~\cite{frank2006three}. In single-particle cryo-EM, a micrograph contains several two-dimensional tomographic projection images, each corresponding to a different copy of the molecule, from an unknown 3-D orientation and placed at an unknown location. Section~\ref{sec:math_model} introduces the formation model of a micrograph in detail. The goal is to recover a 3-D molecular structure from a set of noisy micrographs~\cite{elmlund2015cryogenic, cheng2015primer, sigworth2016principles, singer2020computational, bendory2020single}.

The prevalent cryo-EM computational paradigm splits the reconstruction process into two main stages. The first stage involves identifying and extracting the projection images from the micrographs. This stage is called particle picking, see for example~\cite{wang2016deeppicker, heimowitz2018apple, bepler2019positive, eldar2020klt}. In the second stage, the 3-D structure is reconstructed from the extracted projection images.  Clearly, the quality of the reconstruction depends on the quality of the particle picking stage, which in turn depends heavily on the signal-to-noise ratio (SNR) of the micrograph~\cite{sigworth2004classical}. Therefore, this approach fails when the SNR of the micrograph is very low. In particular, it fails for small molecular structures that induce low SNR  because fewer electrons carry information. The detection threshold has been recognized early on as a central limiting factor by the cryo-EM community; it was suggested that particle picking is impossible for molecules with molecular weight below~$\sim 40 \text{ kDa}$~\cite{henderson1995potential, glaeser1999electron}. Indeed, to date, the vast majority of biomolecules whose structures have been determined using cryo-EM have molecular weights greater than~$100 \text{ kDa}$. Recovering smaller molecular structures is of crucial importance in cryo-EM, and is an active focal point of research endeavors in the field~\cite{wu2012fabs, danev2017expanding, scapin2018cryo, liu20193, zhang2019cryo, wu2020low, yeates2020development, bai2021seeing, wu2021cryo, zheng2022uniform, kimanius2024data, harrison2023review, schwartz2019laser}.

The failure of the current cryo-EM computational paradigm to recover 3-D structures from  low SNR micrographs can be understood through the lens of classical estimation theory. Assume the 3-D volume is represented by~$M$ parameters. Each particle projection is associated with five pose parameters---the 3-D rotation and the 2-D location. Thus, if we wish to jointly estimate the 3-D structure and the pose parameters of the~$T$ projection images, as in older cryo-EM algorithms~\cite{harauz1983direct},  the number of parameters to be estimated is~$M + 5T$, namely, grows linearly with the number of particle projections. In this case, it is well-known that the existence of a consistent estimator is not guaranteed; see for example the celebrated ``Neyman-Scott paradox''~\cite{neyman1948consistent} and the multi-image alignment problem~\cite{aguerrebere2016fundamental}. Current approaches in cryo-EM can be thought of as ``hybrid'' in the sense that they estimate the locations of the particle projections in the particle picking stage (overall~$2T$ parameters), and marginalize over the rotations (as well as over small translations relative to the estimated locations), see for example~\cite{scheres2012relion}. Thus, the number of parameters is~$M+2T$, which still scales linearly with the number of projections. Indeed, as discussed above, this strategy is not consistent when the SNR is very low since particle picking fails. In this paper, we follow~\cite{bendory2023toward} and aim to marginalize over all nuisance variables---the locations and rotations. In this case, the number of parameters to be estimated is fixed, so, given enough data, designing a consistent estimator may be feasible. Therefore, from an estimation theory viewpoint, recovery in low SNR environments (and thus of small molecular structures) is potentially within reach. 

The authors of~\cite{bendory2023toward} proposed to recover the 3-D volume directly from the micrographs using autocorrelation analysis, but their reported reconstructions were limited to low resolution. In this paper, we propose an alternative computational scheme for high-resolution structure reconstruction based on the expectation-maximization (EM) algorithm~\cite{dempster1977maximum}. EM is an algorithm for finding a local maximum of a likelihood function with nuisance variables. It is widely used in many machine learning and statistics tasks, with applications to parameter estimation~\cite{feder1988parameter}, mixture models~\cite{segol2021improved}, deep belief networks~\cite{hinton2006fast}, and independent component analysis~\cite{hyvarinen2013independent}, to name but a few. The EM algorithm was introduced to the cryo-EM community in~\cite{sigworth1998maximum}, and is by now the most popular method for 3-D recovery from picked particles~\cite{scheres2012relion}, where the 3-D rotations, but not the 2-D locations, are treated as nuisance variables.

In order to recover the molecular structure directly from the micrograph, we aim to develop an EM algorithm that marginalizes over both 2-D translations and 3-D rotations. However, as we show in Section~\ref{subsec:approximate_EM_intro}, a direct application of EM is computationally intractable for our model since the number of possible projection locations in the micrograph grows quickly with the micrograph size. Therefore, based on~\cite{kreymer2022approximate, lan2020multi}, we develop an EM algorithm that maximizes an approximation of the likelihood function. The computational complexity of the algorithm is linear in the micrograph size. To further accelerate the algorithm, we apply a stochastic variant of the approximate EM algorithm, which decreases the computational complexity and memory requirements of each iteration (at the potential cost of additional iterations); see Section~\ref{subsec:s-patchEM_algorithm} for further details.

\subsection*{Contributions}

This work introduces a new computational framework for high-resolution recovery of 3-D molecular structures directly from cryo-EM micrographs, without particle picking. Our main contributions are as follows:

\begin{itemize}
    \item We formulate a full 3-D probabilistic model that explicitly marginalizes over both the 3-D rotations and 2-D translations of the projections, resulting in a fixed number of parameters to estimate. This formulation extends the theoretical ideas of~\cite{bendory2023toward} beyond autocorrelation analysis into a likelihood-based estimation framework.
    \item We develop a tractable approximate EM algorithm tailored to the 3-D cryo-EM setting. While conceptually related to 1-D and 2-D approximate EM schemes~\cite{kreymer2022approximate, lan2020multi}, the 3-D case introduces new challenges, including the discretization over $\SO(3)$, the handling of 2-D tomographic projections within the scheme, and the computational scaling with the size of the volume, which require fundamentally different algorithmic and numerical strategies.
    \item We introduce a stochastic variant of the approximate EM algorithm, enabling efficient \rev{processing} on large micrographs and improving scalability to realistic cryo-EM data.
    \item We demonstrate through numerical experiments that our method achieves substantially higher resolution than autocorrelation-based approaches~\cite{bendory2023toward}, validating that direct recovery from micrographs is feasible. 
    %in low SNR regimes.
    \rev{While these experiments are conducted at moderate noise levels where particle picking may still be feasible, they provide the necessary algorithmic foundation for future work on extremely low SNR data where particle picking is impossible.}
\end{itemize}

Taken together, these contributions establish that consistent and high-resolution recovery from micrographs is possible without explicit particle picking, paving the way toward extending cryo-EM to smaller and more flexible molecular structures.

\subsection*{Paper Outline}
The remainder of this paper is organized as follows. Section~\ref{sec:math_model} details the mathematical formation model of a micrograph. Section~\ref{sec:EM} introduces the approximate EM formulation, including the stochastic variant and frequency marching used to accelerate the algorithm. We also discuss the noise level constraints and computational load in Section~\ref{subsec:complexity}. In Section~\ref{sec:results}, we demonstrate that the proposed EM can accurately estimate molecular structures from simulated data, outperforming the autocorrelation analysis of~\cite{bendory2023toward}.  Finally, Section~\ref{sec:conclusions} summarizes our findings and outlines potential strategies for alleviating the computational complexity to enable application to experimental micrographs of extremely low SNR, which is essential for the reconstruction of small molecular structures.

\section{Measurement formation model}
\label{sec:math_model}
Our micrograph formation model follows the  formulation  of~\cite{bendory2023toward}. Let~$f: \mathbb{R}^3 \rightarrow \mathbb{R}$ represent the 3-D electrostatic potential of the molecule to be estimated. We refer to~$f$ as the volume. A 2-D tomographic projection of the volume is a line integral, given by
\begin{equation}
	I_{\omega} (x, y) := PR_\omega f = \int_{-\infty}^{\infty} \left(R_{\omega} f \right) (x, y, z) dz,
\end{equation}
where the operator~$R_{\omega}$ rotates the volume by ~$\omega\in \SO(3)$ and~$P$ is the tomographic projection operator. The micrograph consists of~$T$ tomographic projections, taken from different viewing directions~$\left\{\omega_t\right\}_{t=1}^T \in \SO(3)$, centered at different positions~$\left\{\left(x_t, y_t\right)\right\}_{t=1}^T$,
\begin{equation}
\begin{split}
    	\mathcal{I} (x, y) &= \int_{-\infty}^{\infty} \sum_{t=1}^T \left(R_{\omega_t} f\right)(x - x_t, y - y_t, z) dz + \varepsilon(x,y) \\&= \sum_{t=1}^T \int_{-\infty}^{\infty} \left(R_{\omega_t} f\right)(x - x_t, y - y_t, z) dz + \varepsilon(x,y) \\& = \sum_{t=1}^T I_{\omega_t} (x - x_t, y - y_t) + \varepsilon(x,y),
\end{split}
\end{equation}
where~$\varepsilon(x, y)$ is assumed to be i.i.d.\ white Gaussian noise with zero mean and variance~$\sigma^2$.

\begin{figure}[tbp]
	\subfloat[$\SNR = 1$.]{
		\centering
		\includegraphics[width=0.3\columnwidth, keepaspectratio]{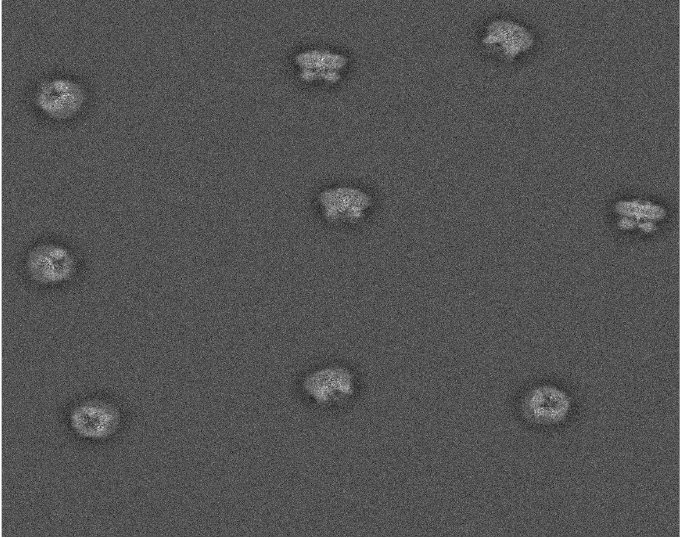}
		\label{fig:SNR1}}
	\hfill
	\subfloat[$\SNR = 1/10$.]{
		\centering
		\includegraphics[width=0.3\columnwidth, keepaspectratio]{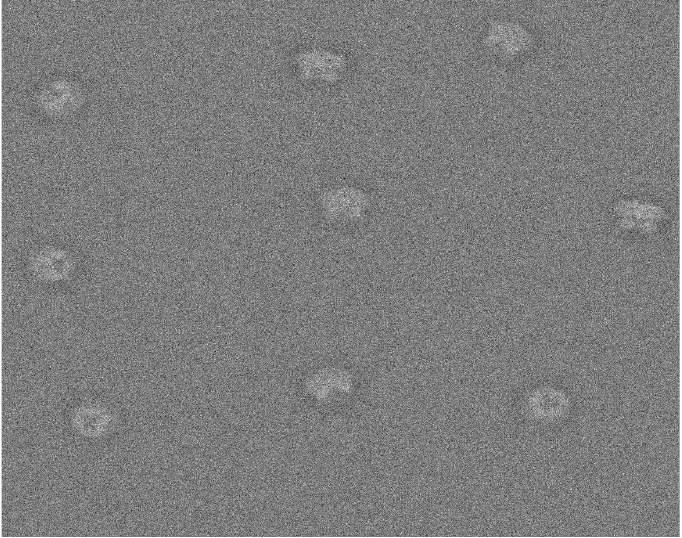}
		\label{fig:SNR01}}
	\hfill
	\subfloat[$\SNR = 1/100$.]{
		\centering
		\includegraphics[width=0.3\columnwidth, keepaspectratio]{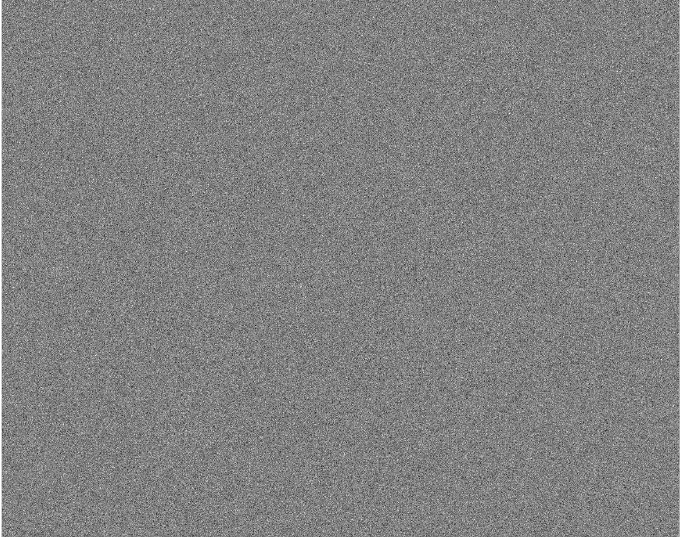}
		\label{fig:SNR001}}
	\caption{Three simulated micrographs at different SNRs. Each measurement contains~$T = 9$ projections of the target volume. We focus on the low SNR regime, where the 2-D locations and 3-D rotations of the projection images cannot be estimated reliably.}
	\label{fig:example_micrographs}
\end{figure}

We  further assume that the micrograph is discretized on a Cartesian grid, the particle projections are centered on the grid, and each projection is of size~$L \times L$ pixels; the projection size,~$L$, is assumed to be known. We denote the indices on the  grid by~$\vec{\ell} = (\ell_x,\ell_y) \in \mathbb{Z}^2$. Thus, our micrograph model~\mbox{$\mathcal{I} \in \mathbb{R}^{N \times N}$} reads
\begin{equation} \label{eq:micrograph}
	\mathcal{I} [\vec{\ell}]  = \sum_{t=1}^T I_{\omega_t} [\ell_x - \ell_{x_t}, \ell_y - \ell_{y_t}]  + \varepsilon[\vec{\ell}].
\end{equation}
The goal is to estimate~$f$ from several micrographs while the rotations, translations, and the number of projections are unknown. Importantly, it is possible to reconstruct the target volume only up to a 3-D rotation, translation, and reflection. Similar mathematical models were thoroughly studied in previous works for one- and two-dimensional setups~\cite{bendory2019multi, lan2020multi, marshall2020image, bendory2023multi, kreymer2022two, shalit2022generalized, kreymer2022approximate}. Figure~\ref{fig:example_micrographs} presents an example of a noisy micrograph~$\mathcal{I}$ at different SNRs, where
\begin{equation}
	\label{eq:SNR}
	\SNR := \frac{\mathbb{E} \left[\lVert I_{\omega_t} \rVert_{\text F}^2 \right]}{L^2 \sigma^2} \approx \frac{\frac{1}{T} \sum_{t=1}^{T} \lVert I_{\omega_t} \rVert_{\text F}^2}{L^2 \sigma^2},
\end{equation}
where~$\lVert \cdot \rVert_{\text F}$ is the Frobenius norm. Section~\ref{sec:conclusions} discusses how to include additional aspects of the cryo-EM reconstruction problem in the proposed technique, such as the effect of the contrast transfer function (CTF)~\cite{erickson1971measurement}, colored noise, and non-uniform distribution of the rotations of the particles over~$\SO(3)$.

Following previous  works~\cite{marshall2020image, bendory2023multi}, we also assume that each translation is separated by at least a full projection length,~$L$, from its neighbors, in both the horizontal and vertical axes. Explicitly,
\begin{equation}
	\label{eq:sep}
	|\ell_{x_t} - \ell_{x_s}| \ge 2L-1 \text{ and } |\ell_{y_t} - \ell_{y_s}| \ge 2L-1, \text{ for all } t \ne s.
\end{equation}
In Section~\ref{subsec:arbitrary_spacing_distribution}, we discuss the implications of mitigating this constraint by allowing the projection images to be arbitrarily close.

\subsection{Volume formation model}
\label{subsec:volume_formation}
The volume~$f$ is smooth and real-valued, and its 3-D Fourier transform,~$\hat{f}$, is finitely expanded by
\begin{equation}
	\label{eq:3-d-fourier-bessel}
	\hat{f}(ck, \theta, \varphi) = \sum_{\ell=0}^{\ell_\text{max}} \sum_{m=-\ell}^{\ell} \sum_{s=1}^{S(\ell)} x_{\ell, m, s} Y_{\ell}^{m} (\theta, \varphi) j_{\ell, s}(k), \quad k \le 1,
\end{equation}
where~$c$ is the bandlimit,~$S(\ell)$ is determined using the Nyquist criterion as described in~\cite{bhamre2017anisotropic},~$j_{\ell, s}$ is the normalized spherical Bessel function, given by
\begin{equation}
	j_{\ell, s}(k) = \frac{4}{\lvert j_{\ell + 1}(u_{\ell, s}) \rvert} j_\ell(u_{\ell, s} k),
\end{equation}
where~$j_\ell$ is the spherical Bessel function of order~$\ell$,~$u_{\ell, s}$ is the~$s$-th positive zero of~$j_\ell$, and~$Y_{\ell}^m$ is the complex spherical harmonic, defined by
\begin{equation}
	Y_{\ell}^m(\theta, \varphi) := \sqrt{\frac{2\ell+1}{4\pi} \frac{(\ell-m)!}{(\ell+m)!}} P_{\ell}^m(\cos \theta) e^{i m \varphi},
\end{equation}
where~$P_{\ell}^m$ are the associated Legendre polynomials with the Condon-Shortley phase. We set~$c = 1/2$ for sampling at the Nyquist rate~\cite{levin20183d}. Under this model, we aim to estimate the expansion coefficients~$x_{\ell, m, s}$ that describe~$f$. Since~$f$ is real-valued,~$\hat{f}$ is conjugate-symmetric and thus the expansion coefficients satisfy $x_{\ell, -m, s} = (-1)^{\ell + m} \overline{x_{\ell, m, s}}$.

Let~$I_{\omega}$ denote the tomographic projection obtained from viewing direction~$\omega \in \SO(3)$. By the Fourier projection-slice theorem (see, e.g.,~\cite{natterer2001mathematics}), its 2-D Fourier transform is given by
\begin{equation}
	\label{eq:projection}
	\widehat{I}_{\omega} (ck, \varphi) = \sum_{\ell, m, m', s} x_{\ell, m, s} D_{m', m}^{\ell}(\omega) Y_{\ell}^{m'}\left(\frac{\pi}{2}, \varphi\right) j_{\ell, s}(k),
\end{equation}
where~$D_{m', m}^{\ell}(\omega)$ is a Wigner-D matrix. Note that the projection images in the micrograph model~\eqref{eq:micrograph} are expressed in the space domain, whereas~\eqref{eq:projection} is expressed in Fourier space. To bridge this gap, we use prolate spheroidal wave functions (PSWFs)~\cite{slepian1964prolate}, as explained next.

\subsection{Expressing the projection image in space domain using the prolate spheroidal wave functions}
The PSWFs are eigenfunctions of the truncated Fourier transform:
\begin{equation}
    \label{eq:PSWF_equation}
    \rev{\alpha_{\nu, n}} \rev{\psi_{\nu, n}}(\textbf{k}) = \int_{\|\textbf{r}\|_2 \le 1} \rev{\psi_{\nu, n}}(\textbf{r}) e^{i c (\textbf{r} \cdot \textbf{k})} d\textbf{r},
\end{equation}
where $c$ is the bandlimit of the eigenfunction $\rev{\psi_{\nu, n}}$, and $\rev{\alpha_{\nu, n}}$ are the associated eigenvalues. The eigenfunctions are orthonormal on the unit disk~$\mathbb{D} := \left\{\textbf{r} \in \mathbb{R}^2, \lVert \textbf{r} \rVert_2 \le 1\right\}$, and they are the most energy concentrated among all~$c$-bandlimited functions on~$\mathbb{D}$, i.e., they satisfy
\begin{equation}
	\rev{\psi_{\nu, n}}(\textbf{r}) = \argmax_{\psi} \frac{\lVert \psi (\textbf{r}) \rVert_{\mathcal{L}^2(\mathbb{D})}}{\lVert \psi (\textbf{r}) \rVert_{\mathcal{L}^2(\mathbb{R}^2)}}.
\end{equation}
Explicitly, the PSWFs are given in polar coordinates by
\begin{equation}
	\label{eq:PSWF_definition}
    \rev{\psi_{\nu, n}}(r, \varphi) = \begin{cases}
    \frac{1}{\sqrt{8\pi}} \rev{\alpha_{\nu, n}} \rev{R_{\nu, n}} (r) e^{i \rev{\nu} \varphi}, &r \le 1, \\
    0, &r > 1,
\end{cases}
\end{equation}
where the range of $\rev{\nu} \in \mathbb{Z}$, $n \in \mathbb{N} \cup \{0\}$ is determined by~\cite[Eq. (8)]{landa2017steerable}, the $\rev{R_{\nu, n}}$ are a family of real, one-dimensional functions, defined explicitly in~\cite[Eq. (66)]{landa2017steerable}, and~$\rev{\alpha_{\nu, n}}$ is the eigenvalue corresponding to the $\rev{(\nu, n)}$-th PSWF~\eqref{eq:PSWF_equation}. From~\eqref{eq:PSWF_definition}, we can also see that the PSWFs are steerable~\cite{freeman1991design}---rotating the image is equivalent to multiplying the eigenfunction $\rev{\psi_{\nu, n}}$ by a phase dependent only on the rotation and the index $\rev{\nu}$. The indices $\rev{\nu}$ and $n$ are referred to, respectively, as the angular index and the radial index.

We may expand the projection~\eqref{eq:projection} in the Fourier domain using the PSWFs:
\begin{equation}
    \widehat{I}_{\omega} (ck, \theta) = \sum_{\rev{\nu}, n} b_{\rev{\nu}, n}(\omega) \rev{\psi_{\nu, n}}(k, \theta).
\end{equation}
The coefficients are given by
\begin{equation}
\begin{split}
    b_{\rev{\nu}, n}(\omega) &= \frac{4}{\sqrt{2\pi}\lvert \rev{\alpha_{\nu, n}}\rvert^2} \int_{0}^{2\pi} \int_{0}^{1} \widehat{I}_{\omega}(ck, \theta) \rev{R_{\nu, n}}(k) e^{-i \rev{\nu} \theta} k dk d\theta \\
    &= \sum_{l, m', m, s} x_{\ell, m, s} \frac{\sqrt{8\pi}}{\rev{\alpha_{\nu, n}}} Y_{\ell}^{m'}\left(\frac{\pi}{2}, 0\right) D_{m', m}^{\ell} (\omega) \\
    &\times \frac{1}{2\pi} \int_{0}^{2\pi} e^{i (m' - \rev{\nu}) \theta} d\theta \int_{0}^{1} j_{\ell, s}(k) \rev{R_{\nu, n}}(k) k dk \\
    &= \sum_{\ell \ge \lvert \rev{\nu} \rvert} \sum_{m, s} x_{\ell, m, s} D_{\rev{\nu}, m}^{\ell}(\omega) \beta_{\ell, s; \rev{\nu}, n},
\end{split}
\end{equation}
where
\begin{equation}
	\beta_{\ell, s; \rev{\nu}, n} := \frac{\sqrt{8\pi}}{\rev{\alpha_{\nu, n}}} Y_{\ell}^{\rev{\nu}}(\pi/2, 0) \int_{0}^{1} j_{\ell, s}(k) \rev{R_{\nu, n}}(k) k dk,
\end{equation}
for $\ell \ge \lvert \rev{\nu} \rvert$, and~$0$ otherwise.

Using the property~\eqref{eq:PSWF_equation}, we can express the projection in space domain as
\begin{equation}
    \label{eq:projection_qSWFs}
    \begin{split}
    I_{\omega}(r, \varphi) &= \left(\frac{c}{2\pi}\right)^2 \sum_{\ell=0}^{\ell_{\text{max}}} \sum_{\rev{\nu}=-\ell}^{\ell} \sum_{m=-\ell}^{\ell} \sum_{n=0}^{n_{\text{max}}(\rev{\nu})} \sum_{s=1}^{S(\ell)} x_{\ell, m, s} \rev{\alpha_{\nu, n}} \beta_{\ell, s; \rev{\nu}, n} D_{\rev{\nu}, m}^{\ell}(\omega) \rev{\psi_{\nu, n}}(r, \varphi) \\
    &:= \sum_{\ell=0}^{\ell_{\text{max}}} \sum_{\rev{\nu}=-\ell}^{\ell} \sum_{m=-\ell}^{\ell} \sum_{n=0}^{n_{\text{max}}(\rev{\nu})} \sum_{s=1}^{S(\ell)} x_{\ell, m, s} \widehat{\beta}_{\ell, s; \rev{\nu}, n} D_{\rev{\nu}, m}^{\ell}(\omega) \rev{\psi_{\nu, n}}(r, \varphi),
    \end{split}
\end{equation}
where~$n_{\text{max}}$ is chosen according to~\cite[Eq. (8)]{landa2017steerable}, and~$\widehat{\beta}_{\ell, s; \rev{\nu}, n} := \left(\frac{c}{2\pi}\right)^2 \rev{\alpha_{\nu, n}} \beta_{\ell, s; \rev{\nu}, n}$.

Using~\eqref{eq:micrograph} and~\eqref{eq:projection_qSWFs}, we can now relate the parameters of the volume directly with the micrograph, bypassing the particle picking stage. In the following section, we present our scheme for estimating the coefficients~$x_{\ell, m, s}$ from a set of noisy micrographs.

\section{An approximate expectation-maximization (EM) algorithm for cryo-EM}
\label{sec:EM}
\subsection{Approximate EM}
\label{subsec:approximate_EM_intro}
The EM algorithm estimates the maximum of a likelihood function by iteratively applying the expectation~(E) and the maximization~(M) steps~\cite{dempster1977maximum}. For the model~\eqref{eq:micrograph}, given a measurement~$\mathcal{I}$, the maximum likelihood estimator (MLE) is the maximizer of~$p(\mathcal{I}; x)$ for the vector of coefficients~$x$~\eqref{eq:3-d-fourier-bessel}. The 2-D translations and 3-D rotations associated with the projection images within the micrograph are treated in our analysis as nuisance variables. In the EM terminology, they are referred to as unobserved or latent variables.

In the E-step of the~$(k+1)$-th iteration of the EM algorithm, one computes $Q(x; x_k)$---the expectation of the complete log-likelihood function, where~$x_k$ is the current estimate of the parameters and the expectation is taken over all admissible configurations of translations and rotations. However, for our model, the number of possible translations in the micrograph grows quickly with the micrograph size,~$N^2$. Consequently, a direct application of EM is computationally intractable. Instead, we follow~\cite{lan2020multi, kreymer2022approximate} and partition the micrograph~$\mathcal{I}$ into~$N_{\text{patches}} = (N / L)^2$ non-overlapping patches~$\{\mathcal{I}_q\}_{q = 1}^{N_{\text{patches}}}$; each patch is of the size of a projection image~$L \times L$. In EM terminology, the patches are referred to as the observed data. The separation condition~\eqref{eq:sep} implies that each patch~$\mathcal{I}_q$ can contain either no projection, a full projection, or part of a projection; overall, there are~$(2L)^2$ possibilities (disregarding rotations). We denote the distribution of translations within a patch by $\rho [\rev{\vec{s}}]$, and require that
\begin{equation}
	\label{eq:rho_constraint}
	\sum_{\rev{\vec{s}} \in \rev{\mathbb{S}}} \rho[\rev{\vec{s}}] = 1, \quad \rho[\rev{\vec{s}}] \ge 0 \text{ for } \rev{\vec{s}} \in \rev{\mathbb{S}},
\end{equation}
where~$\rev{\mathbb{S}} := \left\{0, \ldots, 2L-1\right\}^2$.
Thus, instead of aiming to maximize the likelihood function~$p(\mathcal{I}; x)$, we wish to maximize its surrogate ~$\prod_{q = 1}^{N_{\text{patches}}} p(\mathcal{I}_q; x, \rho)$ using EM. Since the number of possible translations in each patch is independent of the micrograph size, applying EM is now tractable.

\begin{figure}[t]
	\centering
	\subfloat[A projection~\mbox{$I_{\omega} = P R_{\omega} f$} of size~$L = 143 \text{ pixels}$ taken from a viewing direction~$\omega$.]{
		\includegraphics[width=0.28\columnwidth, keepaspectratio]{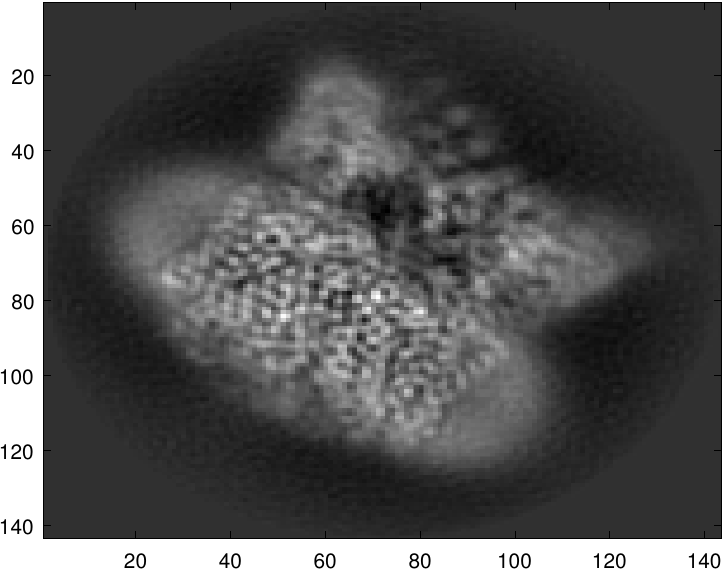}
		\label{fig:F}}
	\hfill
	\subfloat[The padded projection~$Z I_{\omega}$ of size~\mbox{$2L = 286$} pixels.]{
		\includegraphics[width=0.28\columnwidth, keepaspectratio]{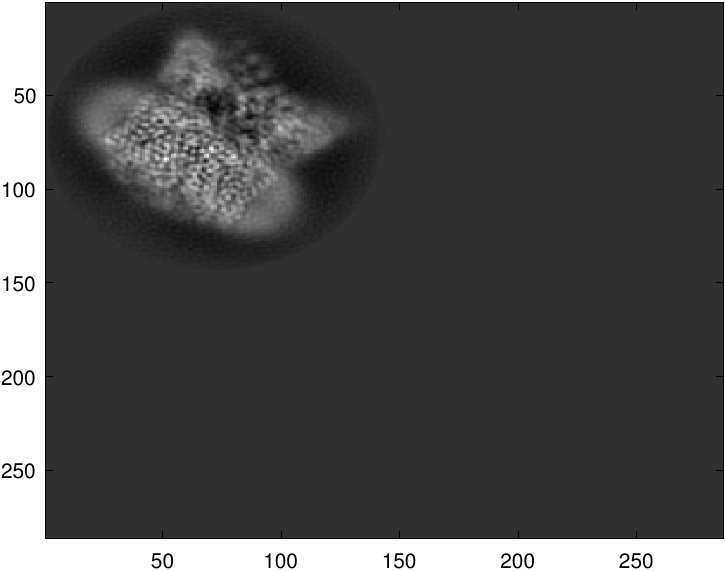}
		\label{fig:ZF}}
	\hfill
	\subfloat[The padded and shifted projection~$T_{{\rev{\vec{s}}}} Z I_{\omega}$ shifted by $\rev{\vec{s}}= (43, 55)$ pixels.]{
		\includegraphics[width=0.28\columnwidth, keepaspectratio]{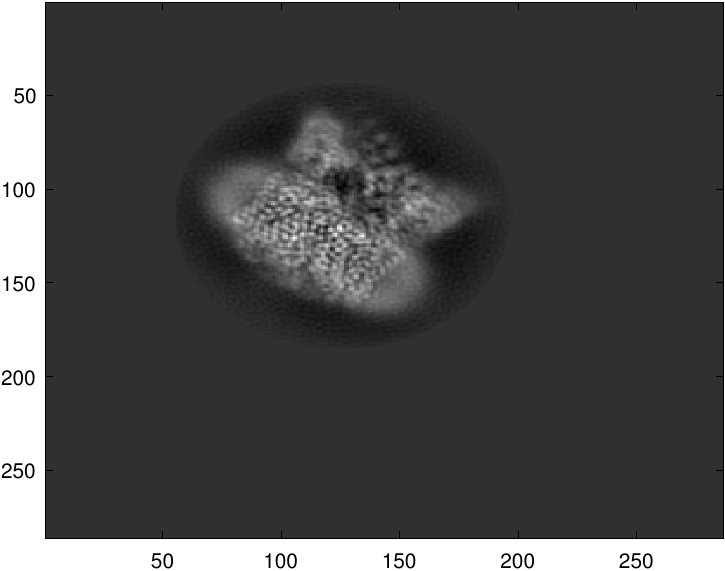}
		\label{fig:TZF}}
	
	\medskip
	
	\subfloat[The cropped image~$C T_{\rev{\vec{s}}} Z I_{\omega}$ of size~$L = 143$ pixels.]{
		\includegraphics[width=0.28\columnwidth, keepaspectratio]{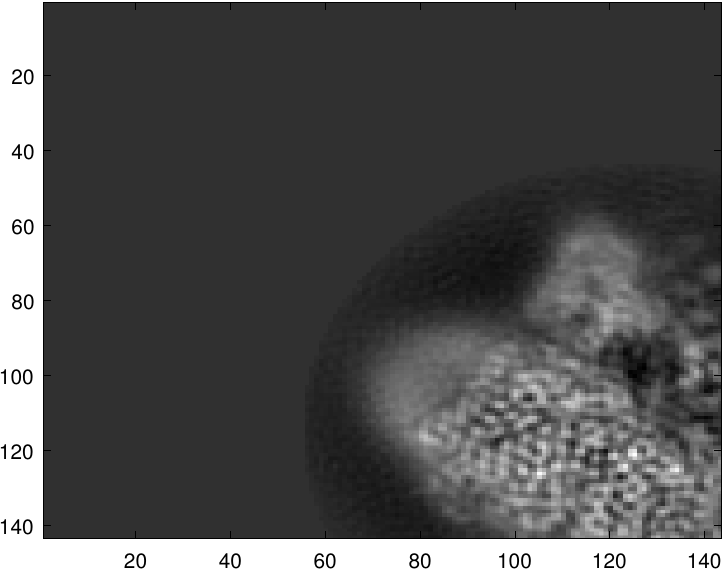}
		\label{fig:CTZF}}
	\quad
	\subfloat[The noisy patch~$\mathcal{I}$ with~$\SNR = 0.5$.]{
		\includegraphics[width=0.28\columnwidth, keepaspectratio]{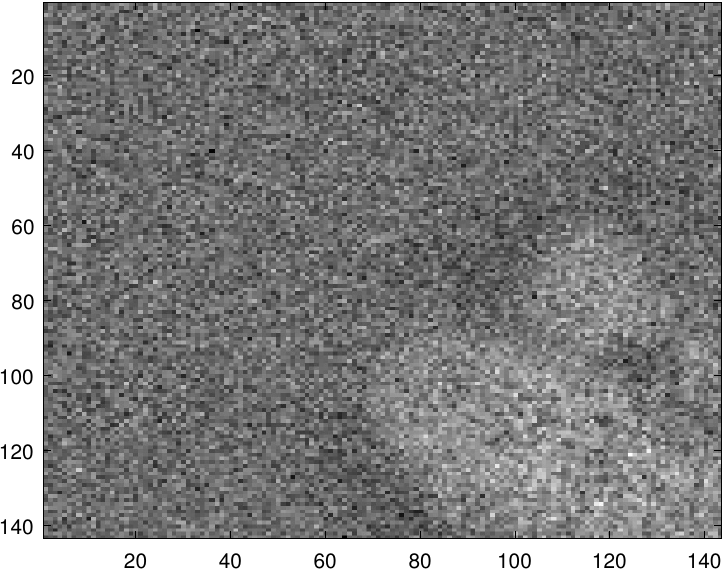}
		\label{fig:CTZF_E}}
	
	\caption{An illustration of the patch generation model described in~\eqref{eq:patch}.}
	\label{fig:patch}
\end{figure}

Specifically, each patch is modeled by
\begin{equation}
	\label{eq:patch}
	\mathcal{I}_q = C T_{\rev{\vec{s}_q}} Z P R_{\omega_q} f + \varepsilon_q, \quad \varepsilon_q \sim \mathcal{N}(0, \sigma^2 I_{L \times L}),
\end{equation}
where the operator~$R_{\omega_q}$ rotates the volume~$f$ by~$\omega_q \in \SO(3)$, and the operator~$P$ projects the volume into 2-D so that~$P R_{\omega_q} f$ is given by~$I_{\omega_q}$~\eqref{eq:projection_qSWFs}. The operator~$Z$ zero-pads~$L$ entries to the right and to the bottom of a projection, and~$T_{\rev{\vec{s}_q}}$ circularly shifts the zero-padded image by $\rev{\vec{s}_q = (s_{x_q}, s_{y_q})\in \mathbb{S}}$ positions, that is,
\begin{equation}
	(T_{\rev{\vec{s}_q}} Z P R_{\omega_q} f )\left[i, j\right] = (Z P R_{\omega_q} f) \left[(i + \rev{s_{x_q}}) \bmod 2L, (j + \rev{s_{y_q}}) \bmod  2L\right].
\end{equation}
The operator~$C$ then crops the first~$L$ entries in the vertical and horizontal axes, and the result is further corrupted by additive white Gaussian noise with zero mean and variance~$\sigma^2$. The generative model of a patch  is illustrated in Figure~\ref{fig:patch}.

Since in the E-step the algorithm assigns probabilities to rotations, the space of rotations must be discretized. We denote the set of~$K$ discrete rotations by~$\Omega^K \subseteq \SO(3)$, such that~$\omega_q \stackrel{\text{i.i.d.\ }}{\sim} \Unif \left\{\Omega^K \right\}$; see Section~\ref{sec:results} for details. Higher~$K$ provides higher accuracy at the cost of running time.

\subsection{EM iterations}
\label{subsec:patchEM_algorithm}
\subsubsection{The E-step}
Given a current estimate of the expansion coefficients~$x_k$ and the distribution of translations~$\rho_k$, in the E-step, our algorithm calculates the expected log-likelihood
\begin{equation}
	Q\left(x, \rho | x_k, \rho_k\right) = \mathbb{E} \left[ \log \mathfrak{L} | \mathcal{I}_1, \ldots, \mathcal{I}_{N_{\text{patches}}}; x_k, \rho_k \right],
\end{equation}
where
\begin{equation}
	\label{eq:approx_likelihood}
	\mathfrak{L} = \prod_{q = 1}^{N_{\text{patches}}} p(\mathcal{I}_q, \rev{\vec{s}_q}, \omega_q;x, \rho),
\end{equation}
which is a surrogate of the computationally intractable complete likelihood function. The expectation is taken over the possible translations and rotations, to achieve
\begin{equation}
	\label{eq:Q}
	Q(x, \rho | x_k, \rho_k) = \sum_{q = 1}^{N_{\text{patches}}} \sum_{\rev{\vec{s}} \in \rev{\mathbb{S}}} \sum_{\omega \in \Omega^K} p(\rev{\vec{s}}, \omega | \mathcal{I}_q; x_k, \rho_k) \log p(\mathcal{I}_q, \rev{\vec{s}}, \omega | x, \rho).
\end{equation}
Applying Bayes' rule, we have that
\begin{equation}
	\label{eq:likelihood_rotations_shifts}
	p(\rev{\vec{s}}, \omega|\mathcal{I}_q; x_k, \rho_k) = \frac{p(\mathcal{I}_q|\rev{\vec{s}}, \omega; x_k, \rho_k) p(\rev{\vec{s}}, \omega|x_k, \rho_k)}{\sum_{\rev{\vec{s}'} \in \rev{\mathbb{S}}} \sum_{\omega' \in \Omega^K} p(\mathcal{I}_q|\rev{\vec{s}'}, \omega'; x_k, \rho_k) p(\rev{\vec{s}'}, \omega'|x_k, \rho_k)},
\end{equation}
which is just the normalized likelihood function
\begin{equation}
	\label{eq:likelihood_qatch}
	p(\mathcal{I}_q|\rev{\vec{s}}, \omega; x_k, \rho_k) \propto \exp \left(- \frac{\|\mathcal{I}_q - C T_{\rev{\vec{s}}} Z P R_{\omega} f\|_{\text F}^2}{2 \sigma^2} \right),
\end{equation}
with the normalization~$\sum_{\rev{\vec{s}} \in \rev{\mathbb{S}}} \sum_{\omega \in \Omega^K} p(\mathcal{I}_q|\rev{\vec{s}}, \omega; x_k, \rho_k) = 1$, weighted by the prior distribution~$p(\rev{\vec{s}}, \omega| x_k, \rho_k) = \frac{\rho_k[\rev{\vec{s}}]}{K}$.

Thus, we can rewrite the expected log-likelihood function~\eqref{eq:Q}, up to a constant, as:
\begin{equation}
	\label{eq:Q_function}
    \begin{split}
	Q(x, \rho|x_k, \rho_k) &= \sum_{q = 1}^{N_{\text{patches}}} \sum_{\rev{\vec{s}} \in \rev{\mathbb{S}}} \sum_{\omega \in \Omega^K} p(\mathcal{I}_q|\rev{\vec{s}}, \omega, x_k) \rho_k[\rev{\vec{s}}] \\ &\times \left(\log p(\mathcal{I}_q|\rev{\vec{s}}, \omega, x) + \log \rho[\rev{\vec{s}}]\right).
    \end{split}
\end{equation}

\subsubsection{The M-step}
The M-step updates~$x$ and~$\rho$ by maximizing~$Q(x, \rho|x_k, \rho_k)$ under the constraint that~$\rho$ is a distribution function:
\begin{equation}
	\label{eq:M_step}
	x_{k+1}, \rho_{k+1} = \argmax_{x, \rho} Q(x, \rho|x_k, \rho_k) \text{ s.t. } \sum_{\rev{\vec{s}} \in \rev{\mathbb{S}}} \rho[\rev{\vec{s}}] = 1, \quad \rho[\rev{\vec{s}}] \ge 0 \text{ for } \rev{\vec{s}} \in \rev{\mathbb{S}}.
\end{equation}

\begin{algorithm}
	\caption{An approximate EM for cryo-EM}
	\label{alg:stochastic_approximate_EM}
	\textbf{Input}: measurement~$\mathcal{I}$ partitioned to~$N_{\text{patches}}$ patches; patch size~$L$; parameter~$K$ (number of discretized rotations); noise variance~$\sigma^2$; initial guesses~$x_0$ and~$\rho_0$; stopping parameter~$\epsilon$; maximal number of iterations~$N_{\text{it}}$; stochastic factor~$0 < S \le 1$.\\
	\textbf{Output}: an estimate of~$x$ and~$\rho$.
	\begin{algorithmic}
		\STATE{Set~$k \rightarrow 0$;}
		\STATE{Set~$\mathfrak{I}_{0; S}: \left \lfloor S N_{\text{patches}} \right \rfloor$ patches, drawn from a uniform distribution;}
		\STATE{Calculate~$p(\mathfrak{I}_{0; S}|\rev{\vec{s}}, \omega; x_0, \rho_0)$ according to~\eqref{eq:likelihood_qatch} for the set~$\mathfrak{I}_{0; S}$, all~$\omega \in \Omega^K$, and all $\rev{\vec{s}} \in \rev{\mathbb{S}}$;}
		\STATE{Calculate~$Q_{0}$ according to~\eqref{eq:Q_function} for patches in~$\mathfrak{I}_{0; S}$, and set~$Q_{-1} \rightarrow -\infty$;}
		\WHILE{$\left(Q_k - Q_{k-1}\right) > \epsilon$ and~$k < N_{\text{it}}$}
		\STATE{Update~$x_{k+1}$ according to~\eqref{eq:update_x};}
		\STATE{Update~$\rho_{k+1}$ according to~\eqref{eq:update_rho};}
		\STATE{Set~$\mathfrak{I}_{k+1; S}: \left \lfloor S N_{\text{patches}} \right \rfloor$ patches, drawn from a uniform distribution;}
		\STATE{Calculate~$p(\mathfrak{I}_{k+1; S}|\rev{\vec{s}}, \omega; x_{k+1}, \rho_{k+1})$ according to~\eqref{eq:likelihood_qatch} for the set~$\mathfrak{I}_{k+1; S}$, all~$\omega \in \Omega^K$, and  all~$\rev{\vec{s}} \in \rev{\mathbb{S}}$;}
		\STATE{Calculate~$Q_{k+1}$ according to~\eqref{eq:Q_function} for patches in~$\mathfrak{I}_{k+1; S}$;} 
        \STATE{Set~$k \rightarrow k + 1$;}
		\ENDWHILE
		\RETURN~$x_k, \rho_k$
	\end{algorithmic}
\end{algorithm}

Since~$Q(x, \rho|x_k, \rho_k)$ is additively separable for~$x$ and~$\rho$, we maximize~$\mathcal{L}(x, \rho, \lambda)$ with respect to~$x$ and~$\rho$ separately. We can express the $x$-dependent part of the expected log-likelihood in a least squares form, namely
\begin{equation}
	\label{eq:Q_ls}
	Q(x|x_k, \rho_k) = - \sum_{q = 1}^{N_{\text{patches}}} \sum_{\rev{\vec{s}} \in \rev{\mathbb{S}}} \sum_{\omega \in \Omega^K} w_k^{q, \rev{\vec{s}}, \omega} \|\mathcal{I}_q - A^{\rev{\vec{s}}, \omega} x\|_{\text F}^2,
\end{equation}
where $w_k^{q, \rev{\vec{s}}, \omega} := p(\mathcal{I}_q|\rev{\vec{s}}, \omega, x_k) \rho_k[\rev{\vec{s}}]$. The operator $A_{\rev{\vec{s}}, \omega}$ can be seen as first projecting the rotated volume, i.e., computing $I_\omega$ as~\eqref{eq:projection_qSWFs}, and then zero-padding, shifting by~$\rev{\vec{s}}$ and cropping back to patch size; see patch model~\eqref{eq:patch} for more details. 

The minimizer of \eqref{eq:Q_ls} is given by the solution to the normal equations of the weighted least-squares:
\begin{equation}
	\label{eq:update_x}
	y_k = C_k x_{k + 1},
\end{equation}
where
\begin{equation}
	\label{eq:C_k}
	C_k := \sum_{q = 1}^{N_{\text{patches}}} \sum_{\rev{\vec{s}} \in \rev{\mathbb{S}}} \sum_{\omega \in \Omega^K} w_k^{q, \rev{\vec{s}}, \omega} {A^{\rev{\vec{s}}, \omega}}^T A^{\rev{\vec{s}}, \omega}
\end{equation}
and
\begin{equation}
	\label{eq:y_k}
	y_k := \sum_{q = 1}^{N_{\text{patches}}} \sum_{\rev{\vec{s}} \in \rev{\mathbb{S}}} \sum_{\omega \in \Omega^K} w_k^{q, \rev{\vec{s}}, \omega} {A^{\rev{\vec{s}}, \omega}}^T \mathcal{I}_q.
\end{equation}

We notice that the data~$\left\{\mathcal{I}_q\right\}_{q = 1}^{N_{\text{patches}}}$ and the projection matrix $A^{\rev{\vec{s}}, \omega}$ can be calculated once at the beginning of the algorithm, as are the multiplication between those terms; see Section~\ref{subsec:complexity} for a discussion about the computational complexity of the EM algorithm. We observe that the system of equations~\eqref{eq:update_x} is well-conditioned in our numerical experiments (see Section~\ref{sec:results}) and is efficiently solved using MATLAB’s Cholesky solver via the \textit{mldivide} function.

The distribution $\rho$ is controlled by the probability that a patch is empty, $\upsilon$; we assume that any other shift within the patch is equally probable. In order to update~$\rho_{k+1}$, we will calculate $\upsilon_{k+1}$ by
\begin{equation}
	\label{eq:update_rho}
	\upsilon_{k+1} = \frac{\sum_{q = 1}^{N_{\text{patches}}} \sum_{\omega \in \Omega^K} p\left(\mathcal{I}_q|\rev{\vec{s}} = \left(L, L\right), \omega, x_k \right)}{\sum_{q = 1}^{N_{\text{patches}}} \sum_{\rev{\vec{s}} \in \rev{\mathbb{S}}} \sum_{\omega \in \Omega^K} p\left(\mathcal{I}_q|\rev{\vec{s}}, \omega, x_k\right)},
\end{equation}
and the updated distribution $\rho_{k+1}$ is given by
\begin{equation}
	\label{eq:rho_k1}
	\rho_{k+1}[\rev{\vec{s}}] = \begin{cases}
    \upsilon_{k+1}, &, \rev{\vec{s}} = \left(L, L\right),\\
    \frac{1 - \upsilon_{k+1}}{\left(2L\right)^2 - 1 - 2  \left(2L - 1\right)}, &\text{otherwise}.
    \end{cases}
\end{equation}

\subsection{Stochastic approximate EM}
\label{subsec:s-patchEM_algorithm}
In order to alleviate the computational burden of the approximate EM scheme (see Section~\ref{subsec:complexity}), we apply a stochastic variant. We focus on incremental EM, first introduced in~\cite{neal1998view}. At each iteration, the incremental EM algorithm applies the E-step to a minibatch of the observed data; the parameters of the problem are updated using the standard M-step. In particular, for the model~\eqref{eq:micrograph}, at each iteration, we choose~$\left \lfloor S N_{\text{patches}} \right \rfloor$ patches drawn uniformly from the set of patches, where~\mbox{$0 < S \le 1$}. We denote the set of~$\left \lfloor S N_{\text{patches}} \right \rfloor$ patches by~$\mathfrak{I}_{k; S}$, where~$k$ is the iteration index. A small~$S$ will result in faster and less memory-consuming iterations, at the possible cost of additional iterations. 

The stochastic approximate EM algorithm is summarized in Algorithm~\ref{alg:stochastic_approximate_EM}.

\subsection{Frequency marching}
\label{subsec:frequency_marching}
As another means to reduce the computational complexity of our scheme, we adopt the frequency marching concept, previously applied to cryo-EM tasks~\cite{barnett2017rapid, scheres2012relion, lan2020multi}. We start our stochastic EM procedure (Algorithm~\ref{alg:stochastic_approximate_EM}) with a low target  frequency (small~$\ell_{\text{max}}$) estimate. When the algorithm is terminated, we use the low-frequency estimate as an initial guess for the  EM procedure with a higher target frequency, and then gradually increase the frequencies. This way, the lion's share of iterations is done over the low-frequency estimates. This is crucial since the computational complexity of the algorithm strongly depends on the frequency~$\ell_{\text{max}}$: see the next section.

\subsection{Complexity analysis}
\label{subsec:complexity}
The computational complexity of the approximate EM algorithm depends mainly on the computational complexity of forming and solving the linear system of equations~\eqref{eq:update_x} at each iteration.

The weights~$w_k^{q, \rev{\vec{s}}, \omega}$ can be computed once at the beginning of each iteration. The computational complexity of computing a single entry of~$p(\mathcal{I}_q | \rev{\vec{s}}, \omega, x_k)$, given by~\eqref{eq:likelihood_qatch}, is~$O\left(L^2\right)$. Recall that we have~$N_{\text{patches}} = (N / L)^2$,~$2L^2$ possible shifts, and~$K$ discrete rotations. By utilizing convolutions, we can sum over all shifts at once, and achieve computational complexity of~$O\left(L^2 \log\left(L^2\right) \left(N / L\right)^2 K \right) = O\left( K N^2 \log\left(L\right) \right)$ for the computation of~$p(\mathcal{I}_q | \rev{\vec{s}}, \omega, x_k)$.
% The total computational complexity of the weights' computation is~$O\left(K N^2 L^2 \log\left(L\right)\right)$.

The operator matrix~$A_{\rev{\vec{s}}, \omega}$ operates on the coefficients~$x_{\ell, m, s}$ (all in all,~$O\left(\ell_{\text{max}}^3\right)$ coefficients), and results in a patch of size~$L^2$. Each entry is the result of summation over the indices~$\rev{\nu}$,~$n$ (see~\eqref{eq:projection_qSWFs}), were each summation is of~$O\left(\ell_{\text{max}}\right)$ operations. Hence, the operator matrix is of size~$L^2 \times O\left(\ell_{\text{max}}^3\right)$, where each entry requires~$O\left(\ell_{\text{max}}^2\right)$ operations. 

Next, for the calculation of the term~$C_k$ (see~\eqref{eq:C_k}), we calculate the product of~$A_{\rev{\vec{s}}, \omega}$ and its transpose for each shift and for each rotation, such that the total computational complexity is~\rev{$O\left(K L^4 \ell_{\text{max}}^6\right)$}. This computation is required \textit{only once}, at the beginning of the algorithm. The computational complexity of calculating the term~$C_k$ is~$O\left( N_{\text{patches}} L^2 K\right) = O\left(N^2 K\right)$ further computations.

For the calculation of the term~$y_k$ (see~\eqref{eq:y_k}), we calculate the product of~$A_{\rev{\vec{s}}, \omega}$ with the~$N_{\text{patches}}$ patches. This computation requires~$O\left( \rev{K} \ell_{\text{max}}^3 L^2 \left(N / L\right)^2 \rev{L^2}\right) = O\left( \rev{K} \ell_{\text{max}}^3 N^2 \rev{L^2} \right)$ further computations, and also done \textit{only once}. Therefore, the computational complexity of calculating the term~$y_k$ is~$O\left( N_{\text{patches}} L^2 K \rev{\ell_{\text{max}}^3}\right) = O\left(N^2 K \rev{\ell_{\text{max}}^3}\right)$ further computations.

Finally, the computational complexity of solving the linear system of~$O \left( \ell_{\text{max}}^3\right)$ equations~\eqref{eq:update_x} is of computational complexity of~$O \left( \ell_{\text{max}}^9 \right)$. We divide the computational complexity analysis of the approximate EM algorithm (Algorithm~\ref{alg:stochastic_approximate_EM}) into two parts: computations required only once, at the beginning of the algorithm, and computations required at each EM iteration, with a total of~$N_{\text{iters}}$ iterations:
\begin{equation}
	O\left(\rev{K L^4\ell_{\text{max}}^{6}}  + \rev{K\ell_{\text{max}}^3 N^2 L^2} + N_{\text{iters}} \left( K N^2 \rev{\ell_{\text{max}}^3} + \ell_{\text{max}}^9\right) \right).
\end{equation}

Since we aim to estimate small molecular structures,~$L^3$  (the dimension of the volume), and~$\ell_{\text{max}}$ (the maximum frequency in the expansion~\eqref{eq:3-d-fourier-bessel}) are expected to be rather small. However,~$N^2$---the total number of pixels in all micrographs---is expected to grow as the SNR decreases, and so is the number of iterations~$N_{\text{iters}}$. Therefore, we expect that the computational complexity of the approximate EM algorithm will be governed by~$O \left( N_{\text{iters}} N^2 K \rev{\ell_{\text{max}}^3} \right)$. Thus, the running time is linear in the number of pixels in the micrograph and the number of rotations considered by the EM framework. The latter presents an accuracy vs. running time trade-off~\cite{kreymer2022approximate}.
In the future, we intend to include more sophisticated and hierarchical techniques to control the sampling of the rotation space, similar to existing methods in cryo-EM software~\cite{scheres2012relion, punjani2017cryosparc}. The benefit of the frequency marching procedure (see Section~\ref{subsec:frequency_marching}) is clear from our analysis---the computational complexity depends polynomially on~$\ell_{\text{max}}$.

When using the stochastic variant of our approximate EM (see Algorithm~\ref{alg:stochastic_approximate_EM}), the computational complexity of the algorithm is~$O \left( S \tilde{N}_{\text{iters}} N^2 K \rev{\ell_{\text{max}}^3} \right)$, where~$\tilde{N}_{\text{iters}}$ is the number of iterations. The computational complexity scales linearly with the stochastic factor~$0 < S \le 1$. However, the number of iterations~$\tilde{N}_{\text{iters}}$ is expected to increase as we process fewer patches per iteration.

Practical considerations hinder us from achieving the aforementioned computational complexity. In practice, the quantities that we can precompute, i.e., the projection operator and its products with its transpose and with the patches, cannot be stored in memory on our machine (see Section~\ref{sec:results} for more details on the machine's specifications). Thus, some precomputation is done in practice at each iteration, and memory storage is overridden by computations in sequence to reduce memory usage.

\section{Numerical results}
\label{sec:results}
The code to reproduce all numerical experiments is publicly available at \url{https://github.com/krshay/Approx-EM-cryo-EM}. 

In this section, we present numerical results for the approximate EM recovery algorithm described in Section~\ref{sec:EM}. The micrographs for the experiments were generated as follows. We sample rotation matrices from~$\SO(3)$ uniformly at random as described in~\cite{shoemake1992uniform}. Given a volume and sampled rotation matrices, we generate projections of the volume corresponding to the rotation matrices using the ASPIRE software package~\cite{garrett_wright_2023_7510635}. The projections are then placed in the measurement one by one; for each added projection, it is verified that the separation condition~\eqref{eq:sep} is not violated. The number of projections in the measurement,~$T$, is determined by the required density~$\gamma = T \frac{N^2}{\tilde{L}^2}$, where~$\tilde{L}^2$ is the size of a projection within the micrograph. Ultimately, the micrograph is corrupted by i.i.d.\ Gaussian noise with zero mean and variance corresponding to the desired SNR.

\begin{figure}[t]
\centering
\subfloat[Volume reconstructions: Left: the initial guess; middle: the estimate up to~$\ell = 14$; right: the ground truth volume.]{
\includegraphics[width=0.49\columnwidth, keepaspectratio]{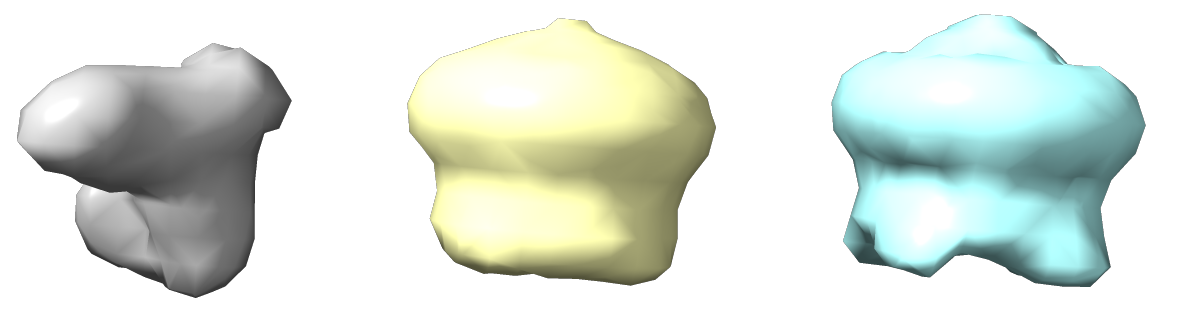}
\label{fig:TRPV1_comp_1}} \vfill
\subfloat[The FSC curves of the volumes in panel (A), with respect to the ground truth.]{
\includegraphics[width=0.50\columnwidth, keepaspectratio]{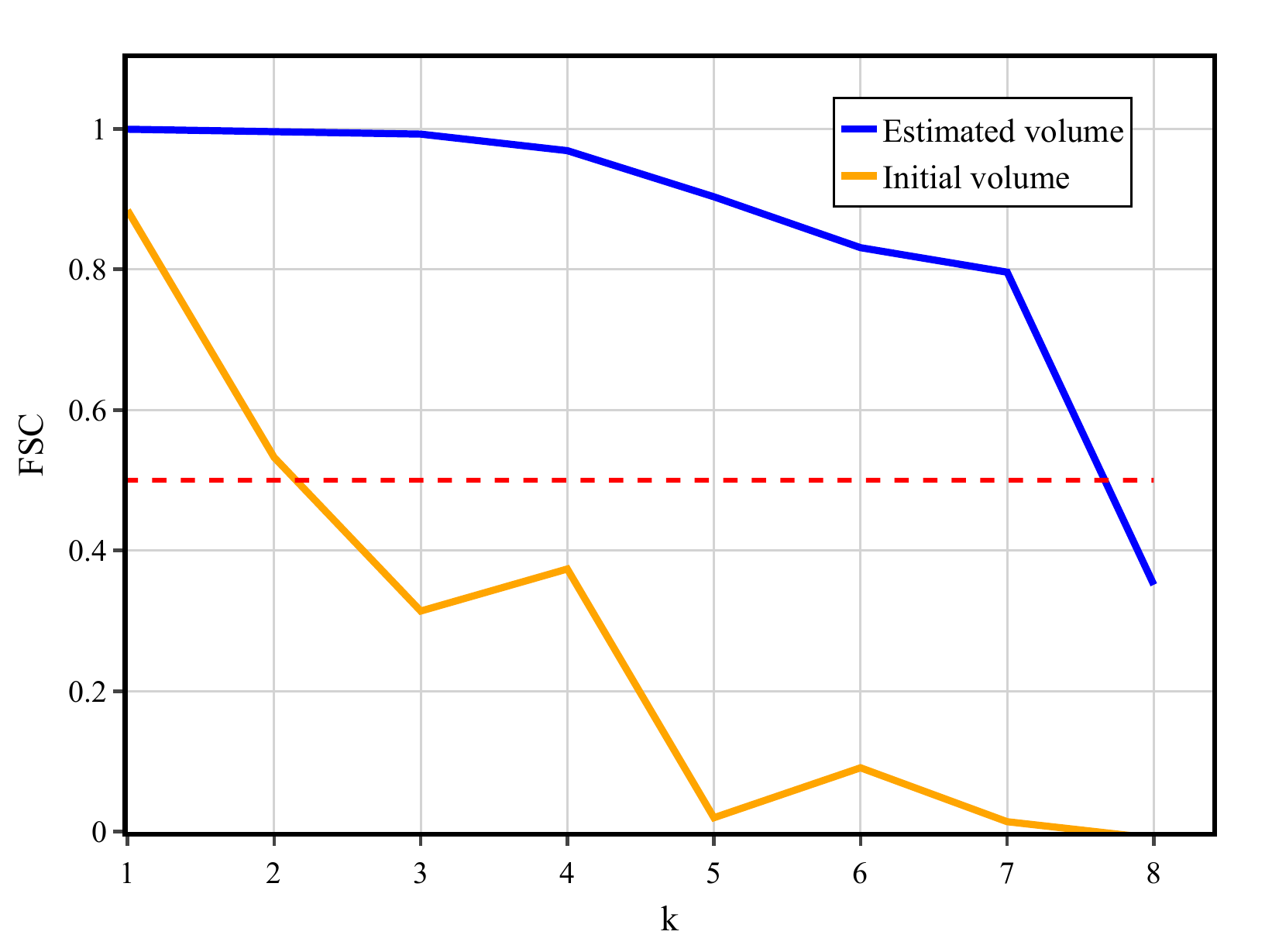} \label{fig:TRPV1_FSC_1}}
\caption{Results for estimating the TRPV1 structure directly from a micrograph. The micrograph was generated from volumes of original size, and then downsampled such that each projection is of size~$17^2$ (Method 1).
\label{fig:TRPV1_results_1}}
\end{figure}

\begin{figure}[t]
\centering
	\subfloat[Volume reconstructions: Left: the initial guess; middle: the estimate up to~$\ell = 14$; right: the ground truth volume.]{
		\includegraphics[width=0.49\columnwidth, keepaspectratio]{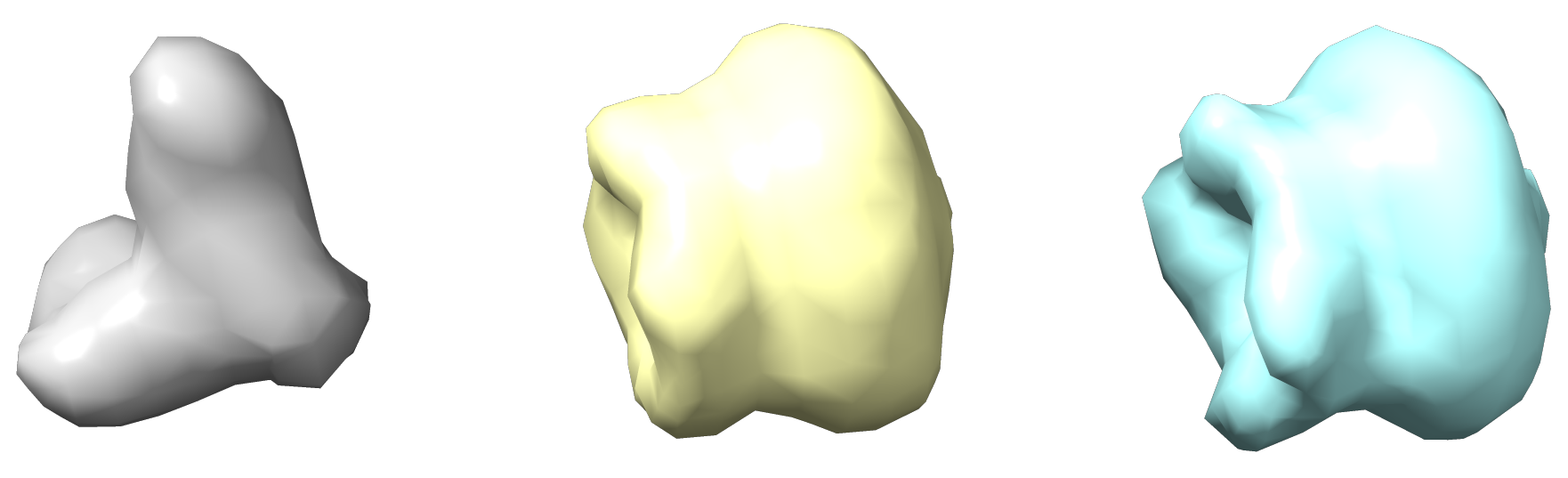}	
		\label{fig:TRPV1_comp_2}} \vfill
	\subfloat[The FSC curves of the volumes in panel (A), with respect to the ground truth.]{
		\includegraphics[width=0.50\columnwidth, keepaspectratio]{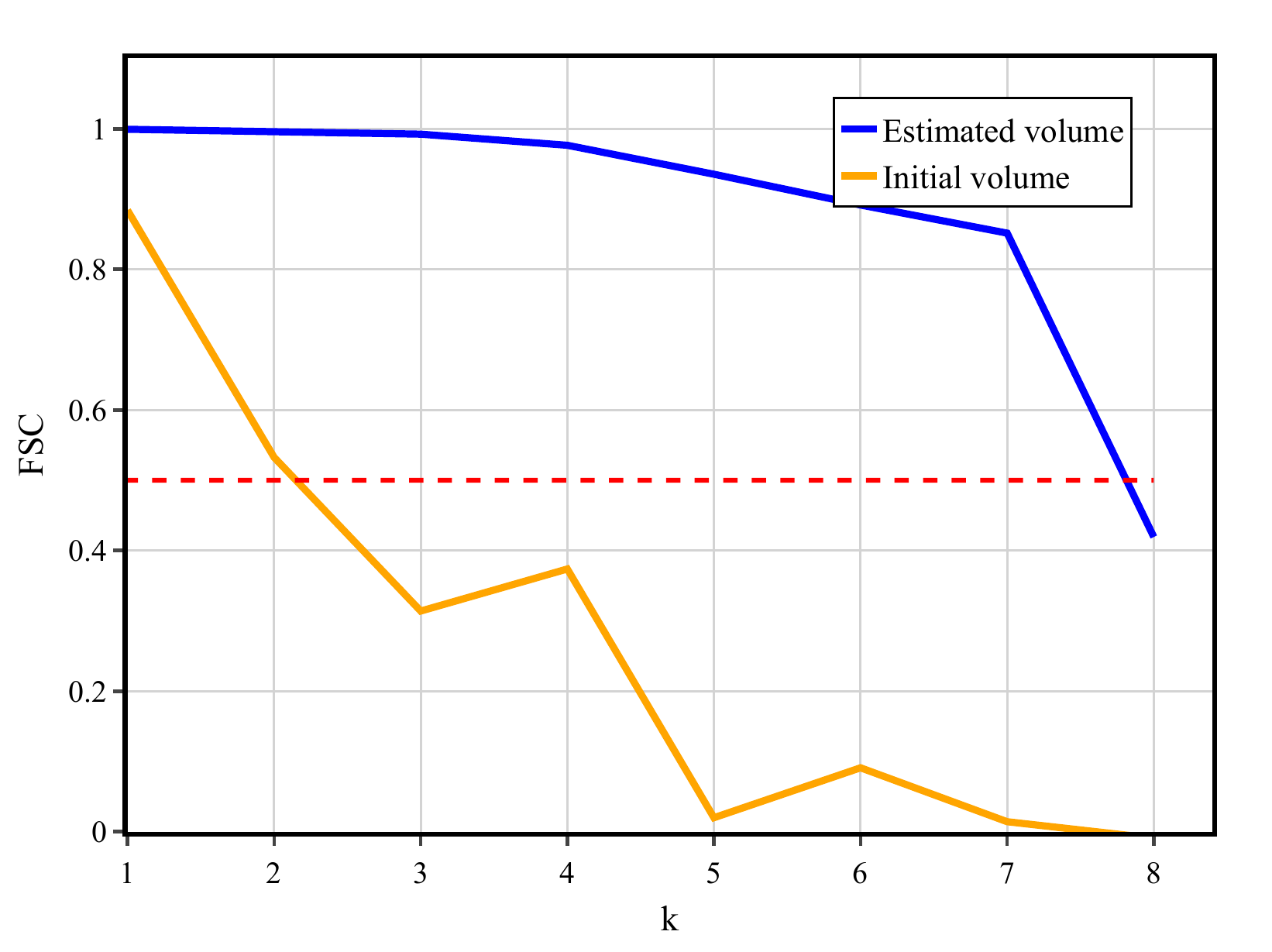}
		\label{fig:TRPV1_FSC_2}}
	\caption{Results for estimating the TRPV1 structure directly from a micrograph. The micrograph was generated from volumes downsampled to size~$17^3$ (Method 2).
\label{fig:TRPV1_results_2}}
\end{figure}

\begin{figure}[ht]
    \centering
    \includegraphics[width=0.49\linewidth]{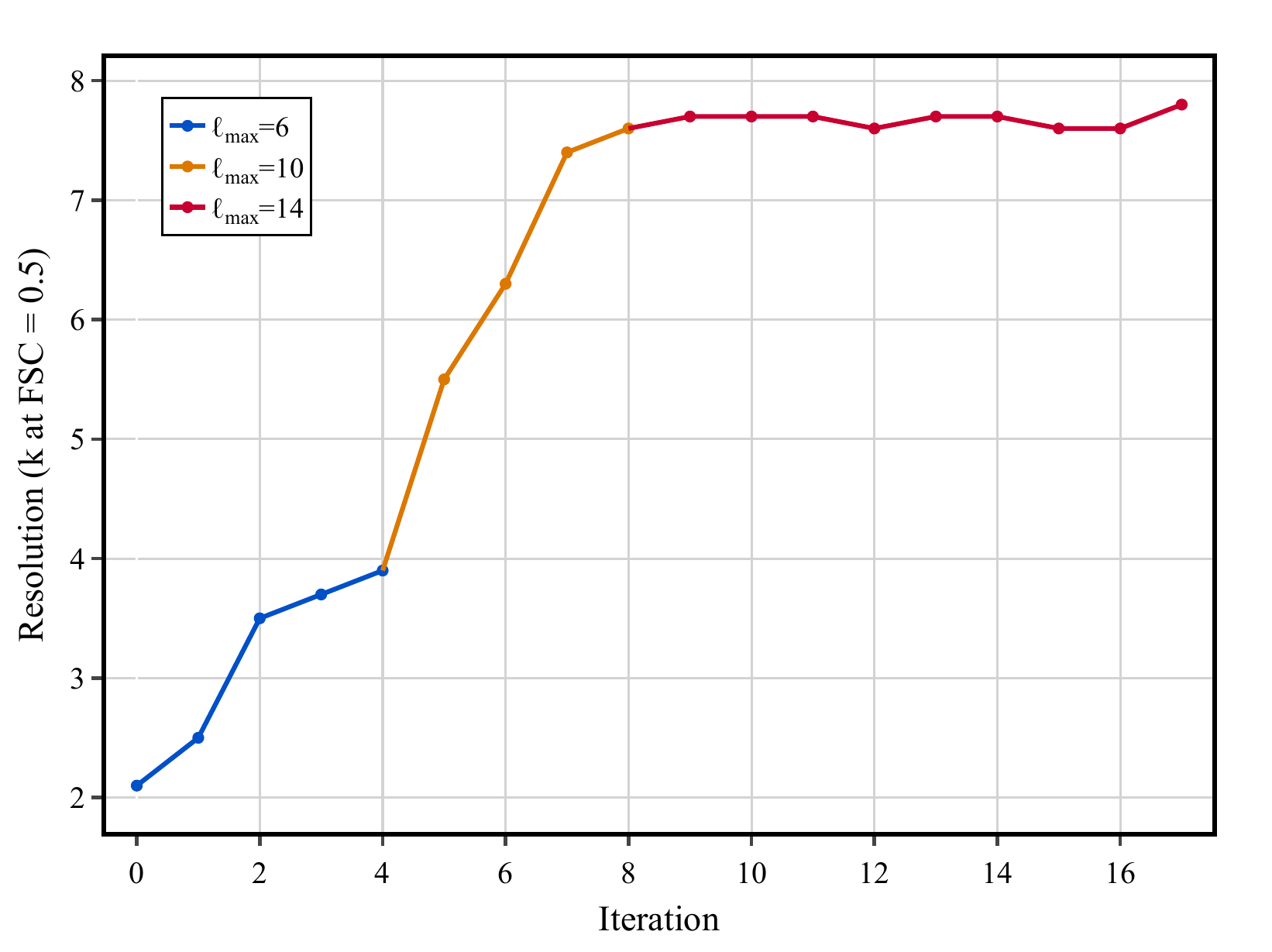}
    \caption{Resolution progression across iterations for the estimation of the TRPV1 volume using Method 2. Resolution is defined as~$\rev{k}$ at which the FSC with the ground truth crosses~$0.5$. Colors indicate the maximal spherical harmonic frequency: blue~($\ell_{\text{max}}=6$), orange~($\ell_{\text{max}}=10$), red~($\ell_{\text{max}}=14$). Higher values of~$\rev{k}$ correspond to finer resolved detail.}
    \label{fig:resolutions}
\end{figure} 

We present reconstructions from micrographs generated in two different ways:
\begin{itemize}
	\item \textbf{Method 1 for micrograph generation}: we generate the micrograph from volumes of the original size. Due to computational constraints, we downsample the micrograph by a factor of~$\tilde{L} / L$, where~$\tilde{L}^3$ is the original size of the volume, and~$L^2$ is the required projection size for computations; we aim to estimate a downsampled volume of size~$L^3$. Importantly, we assume that the downsampled target volumes are bandlimited (as assumed in Section~\ref{subsec:volume_formation}), but do not force it upon them. That is to say, this approximation error bounds the volume recovery error. This method imitates the procedure one will perform on experimental data sets. We note that downsampling improves the SNR because the spectra of the volumes decay faster than the noise spectrum.
	
	\item \textbf{Method 2 for micrograph generation}: we generate the micrograph from volumes downsampled to size~$L^3$. We expect this method to outperform the first (more realistic) method. We note that this micrograph generation scheme is not exactly aligned with our mathematical model of the micrograph generation (see Section~\ref{sec:math_model}), since the target volume is not expanded using the expansion~\eqref{eq:3-d-fourier-bessel}; however, we still expect this method to outperform the first (more realistic) method.
\end{itemize}

We follow the standard convention in the cryo-EM literature and measure the accuracy of the reconstruction using the Fourier shell correlation (FSC) metric. The FSC is calculated by correlating the 3-D Fourier components of two volumes (the ground truth~$f_\text{true}$ and the estimation~$f_\text{est}$) and summing over spherical shells in Fourier space:
\begin{equation}
	\text{FSC} (\rev{k}) = \frac{\sum_{\rev{k}_i \in S_\rev{k}} \hat{f}_\text{true}(\rev{k}_i) \hat{f}^*_\text{est}(\rev{k}_i)}{\sqrt{\sum_{\rev{k}_i \in S_\rev{k}} \lvert \hat{f}_\text{true}(\rev{k}_i)|^2 \sum_{\rev{k}_i \in S_\rev{k}} |\hat{f}_\text{est}(\rev{k}_i)|^2}},
\end{equation}
where~$S_\rev{k}$ is the spherical shell of radius~$\rev{k}$. We use the~$0.5$ resolution cutoff: the resolution is determined as the shell index where the FSC curve drops below 0.5.

\subsection{Volume reconstructions}
\label{subsec:volume_reconstructions}
We present reconstructions of two volumes. All reconstructions were achieved by applying the stochastic approximate EM algorithm (Algorithm~\ref{alg:stochastic_approximate_EM}) with frequency marching. In all experiments, we used a micrograph of size~$N^2 = 1{,}003^2 \text{ pixels}$, with~$T = 1{,}392$ total projections. Each experiment consists of~$N_{\text{patches}} = 3{,}481 \text{ patches}$. We ran 5 iterations with~$\ell_{\text{max}} = 6$, a stochastic factor of~$S = 1$, and~$K = 3{,}392$ discrete rotations; 5 iterations with~$\ell_{\text{max}} = 10$, a stochastic factor of~$S = 0.5$ and~$K = 3{,}392$ discrete rotations; and 10 iterations with~$\ell_{\text{max}} = 14$, a stochastic factor of~$S = 0.25$ and~$K = 1{,}376$ discrete rotations. The experiments were performed on a machine with 96 cores of Intel(R) Xeon(R) Gold 6252 CPU @ 2.10GHz with 1.51 TB of RAM, and took less than approximately 6 hours per EM iteration with~$\ell_{\text{max}} = 6$, approximately 12 hours per iteration with $\ell_{\text{max}} = 10$, and approximately 12 hours per iteration with $\ell_{\text{max}} = 14$. The molecular visualizations were produced using UCSF Chimera~\cite{pettersen2004ucsf}.

\subsubsection[The TRPV1 structure]{The TRPV1 structure~\cite{gao2016trpv1}}
The volume is available at the Electron Microscopy Data Bank (EMDB) as \mbox{EMD-8117}\footnote{\url{https://www.ebi.ac.uk/emdb/}}. The true structure is of size~$\tilde{L}^3 = 192^3 \text{ voxels}$. The initial guess for the algorithm was generated from volume reconstructions from the AlphaFold Protein Structure Database (AFDB)\footnote{\url{https://alphafold.ebi.ac.uk}}~\cite{varadi2024alphafold, varadi2022alphafold}, downsampled to the size of the target volume. We used entry~\mbox{AF-Q8NER1-F1} of AFDB for the initial guess. 

First, we use Method 1 to generate the micrograph, i.e., we generate the micrograph with the true~$192^3 \text{ voxels}$ volume, and then downsample the micrograph such that each projection is of size~$17 \times 17 \text{ pixels}$.  For this experiment, the micrograph was simulated with~$\SNR = 0.13$.  A visual comparison between the true and the estimated volume is presented in Figure~\ref{fig:TRPV1_comp_1}, and the FSC curve is given in Figure~\ref{fig:TRPV1_FSC_1}. Using Method 2 to generate the micrograph, we have generated the micrograph with~$\SNR = 6.2$, and downsampled the volume to size~$L^3 = 17^3 \text{ voxels}$ before micrograph generation. A visual comparison is presented in Figure~\ref{fig:TRPV1_comp_2}, and the FSC curve is provided in Figure~\ref{fig:TRPV1_FSC_2}.

Remarkably, we achieve accurate estimates of the downsampled TRPV1 structure from micrographs generated using both methods. \rev{The resolution (measured by the FSC = 0.5 threshold) improves consistently from iteration 0 (AlphaFold initialization) through the final iteration, as can be see in Figure~\ref{fig:resolutions}. The estimates improve rapidly in the first few iterations and then stabilize, indicating convergence within the reported number of iterations.

}

\begin{figure}[t]
\centering
	\subfloat[Volume reconstructions: Left: the initial guess; middle: the estimate up to~$\ell = 14$; right: the ground truth volume.]{
		\includegraphics[width=0.49\columnwidth, keepaspectratio]{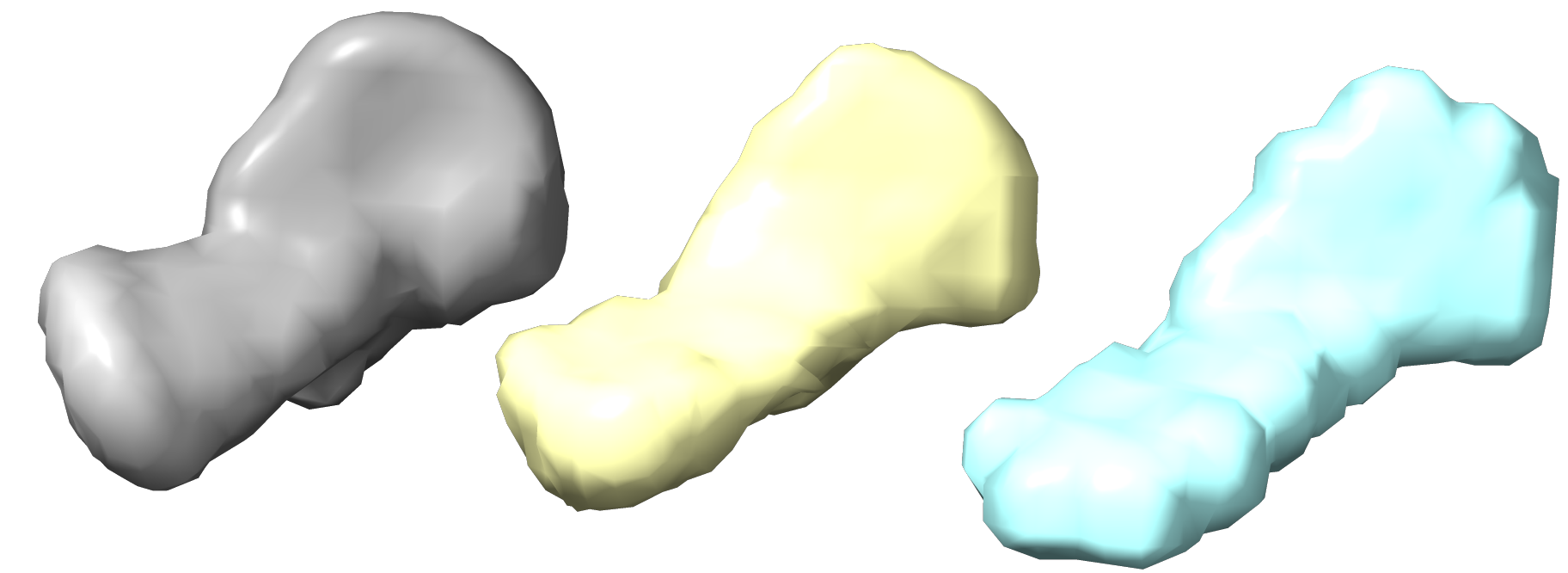}	
		\label{fig:geom_comp_1}} \vfill
	\subfloat[The FSC curves of the volumes in panel (A), with respect to the ground truth.]{
		\includegraphics[width=0.50\columnwidth, keepaspectratio]{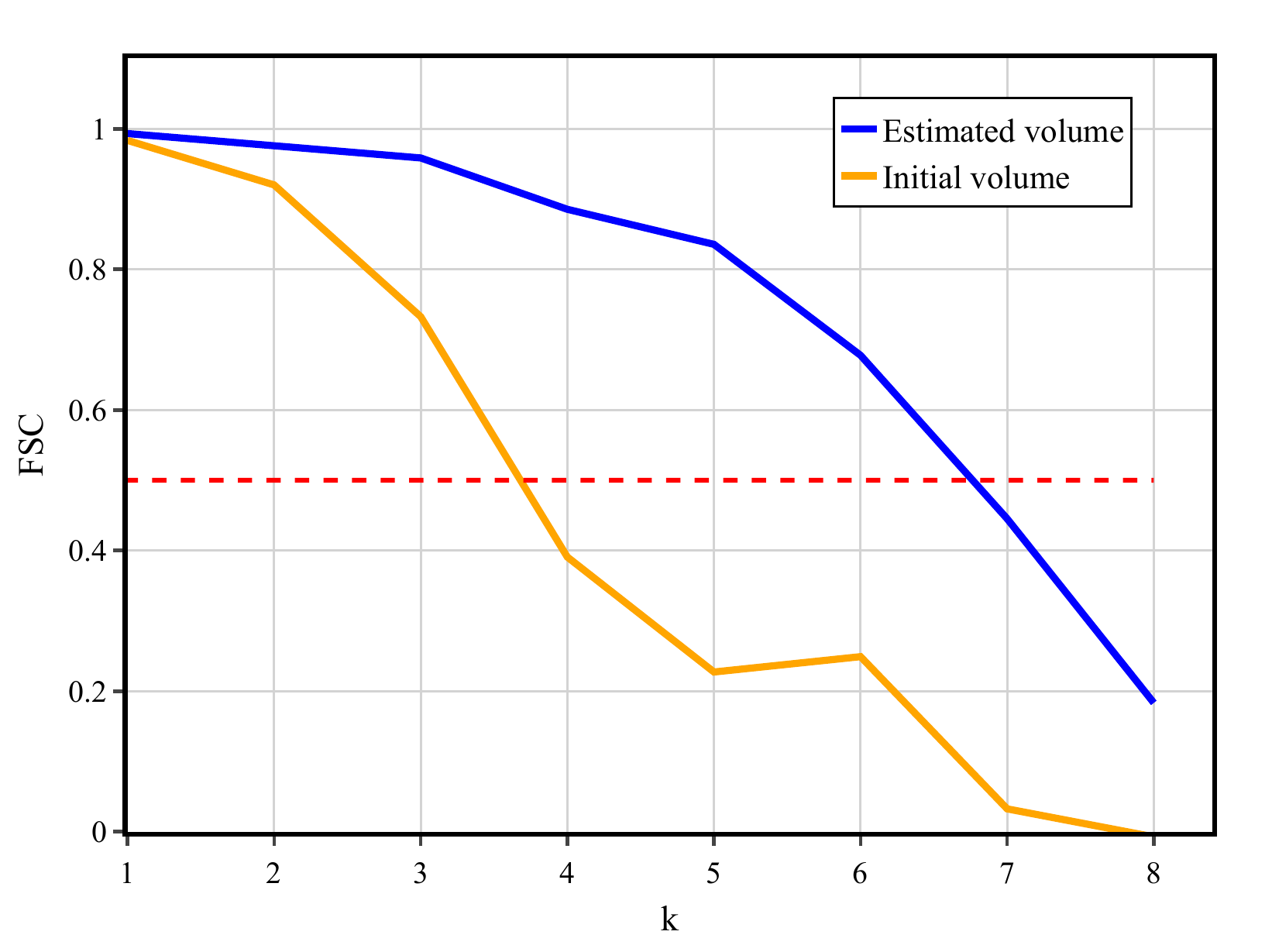}
		\label{fig:geom_FSC_1}}
	\caption{Results for estimating the molecular conformation from GEOM dataset directly from a micrograph, from the first initial guess.
		\label{fig:geom_results_1}}
\end{figure}

\begin{figure}[t]
\centering
	\subfloat[Volume reconstructions: Left: the initial guess; middle: the estimate up to~$\ell = 14$; right: the ground truth volume.]{\includegraphics[width=0.49\columnwidth, keepaspectratio]{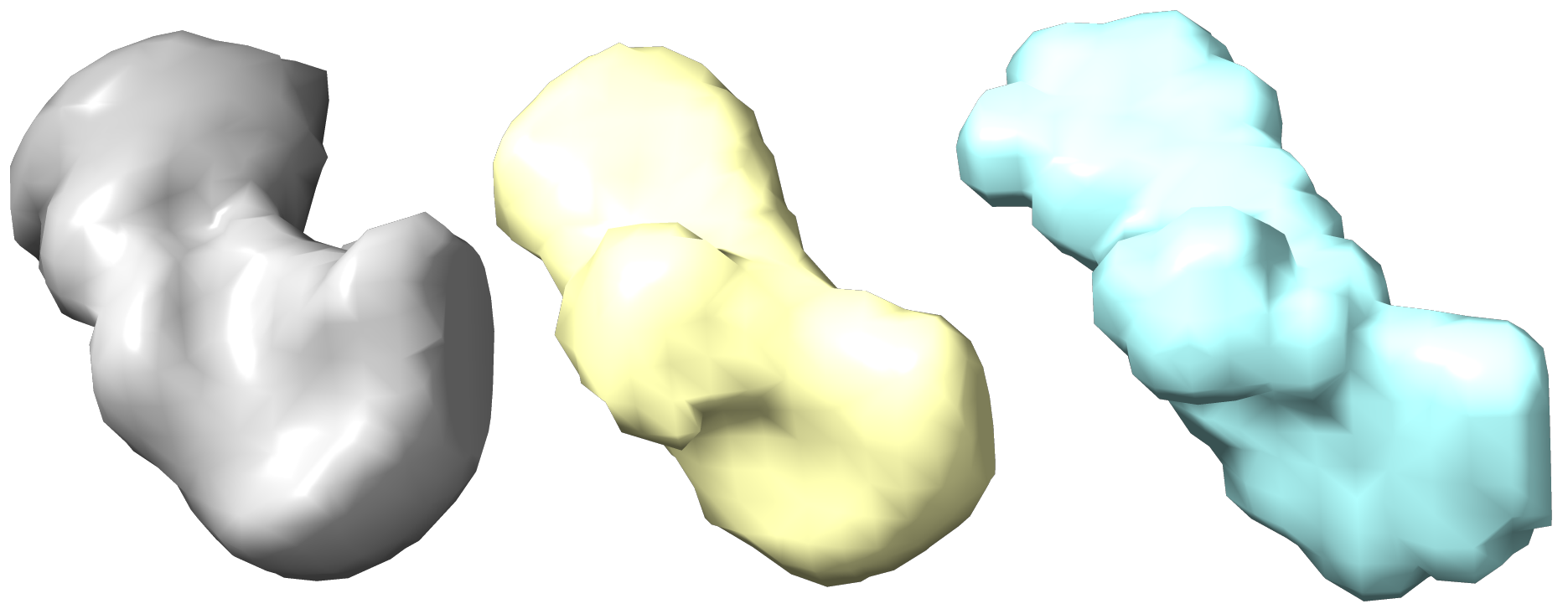}	
		\label{fig:geom_comp_2}} \vfill
	\subfloat[The FSC curves of the volumes in panel (A), with respect to the ground truth.]{
		\includegraphics[width=0.50\columnwidth, keepaspectratio]{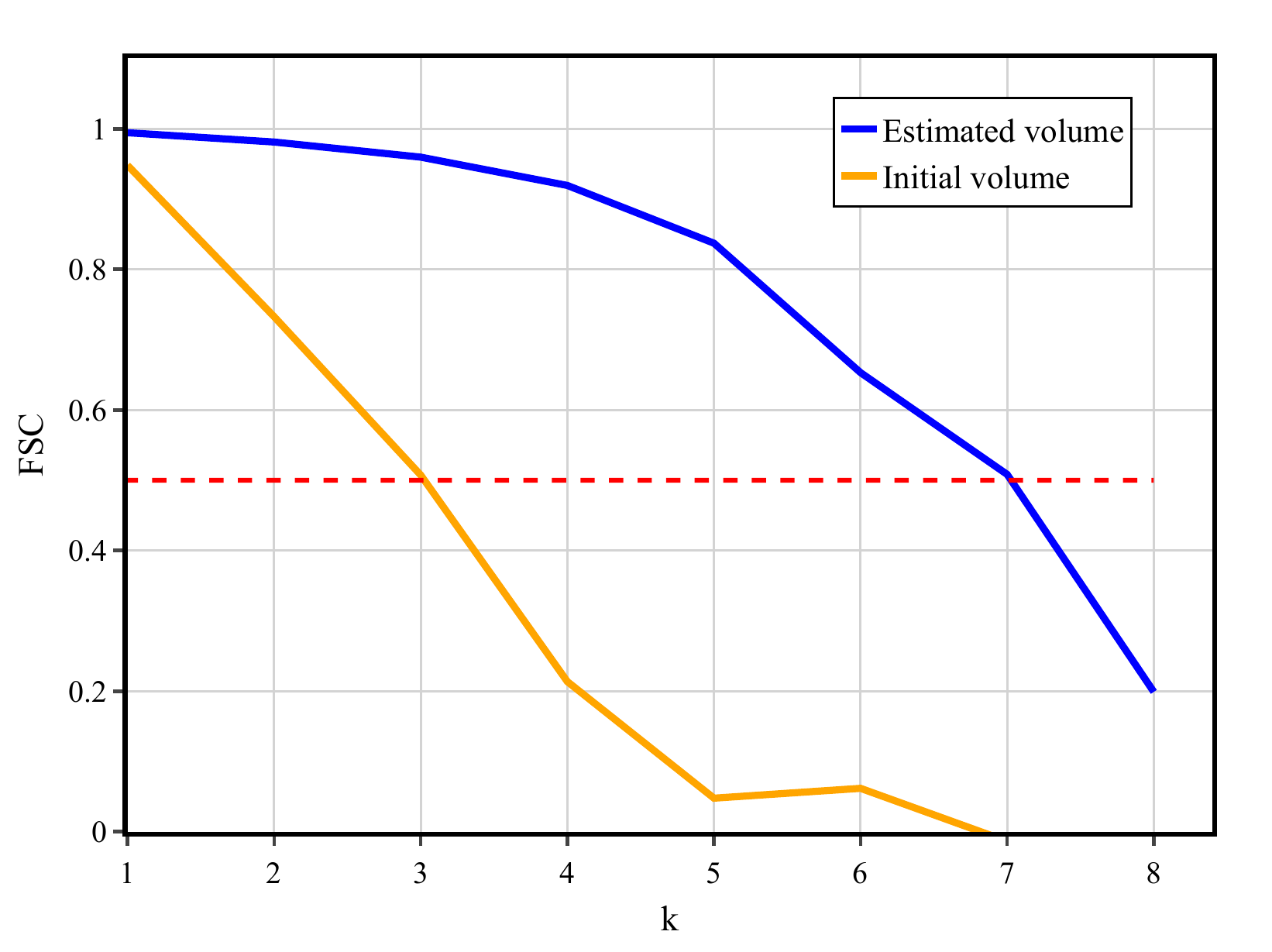}
		\label{fig:geom_FSC_2}}
	\caption{Results for estimating the molecular conformation from GEOM dataset directly from a micrograph, from the second initial guess.
		\label{fig:geom_results_2}}
\end{figure}

\subsubsection[A molecular conformation from the Geometric Ensemble Of Molecules (GEOM)]{A molecular conformation from the Geometric Ensemble of Molecules (GEOM)~\cite{axelrod2022geom}}
GEOM is a dataset with 37 million molecular conformations annotated by energy and statistical weight for over 450,000 molecules. We used the paper's accompanying code to generate volumes of size~$L^3 = 17^3$, and used two conformations of the same molecule as initial guesses to the algorithm.

We have generated the micrographs with~$\SNR = 6.2$. A visual comparison between the true and the estimated volumes is provided in Figures~\ref{fig:geom_comp_1} and~\ref{fig:geom_comp_2}, and the FSC curve is presented in Figures~\ref{fig:geom_FSC_1} and~\ref{fig:geom_FSC_2}. The results demonstrate successful reconstructions, though they also reveal a noticeable dependence on the choice of initial volume.

\rev{
\begin{figure}[t]
\centering
	\subfloat[Volume reconstructions: Gray: the initial guess; yellow: the estimate up to~$\ell = 14$; turquoise: the ground truth volume; purple: the downsampled estimate of the volume using autocorrelation analysis, expanded up to~$\ell_{\text{max}} = 2$, as was estimated in~\cite{bendory2023toward}. The estimation in~\cite{bendory2023toward} was done from clean autocorrelations, corresponding to~$N \rightarrow \infty$.]{
		\includegraphics[width=0.49\columnwidth, keepaspectratio]{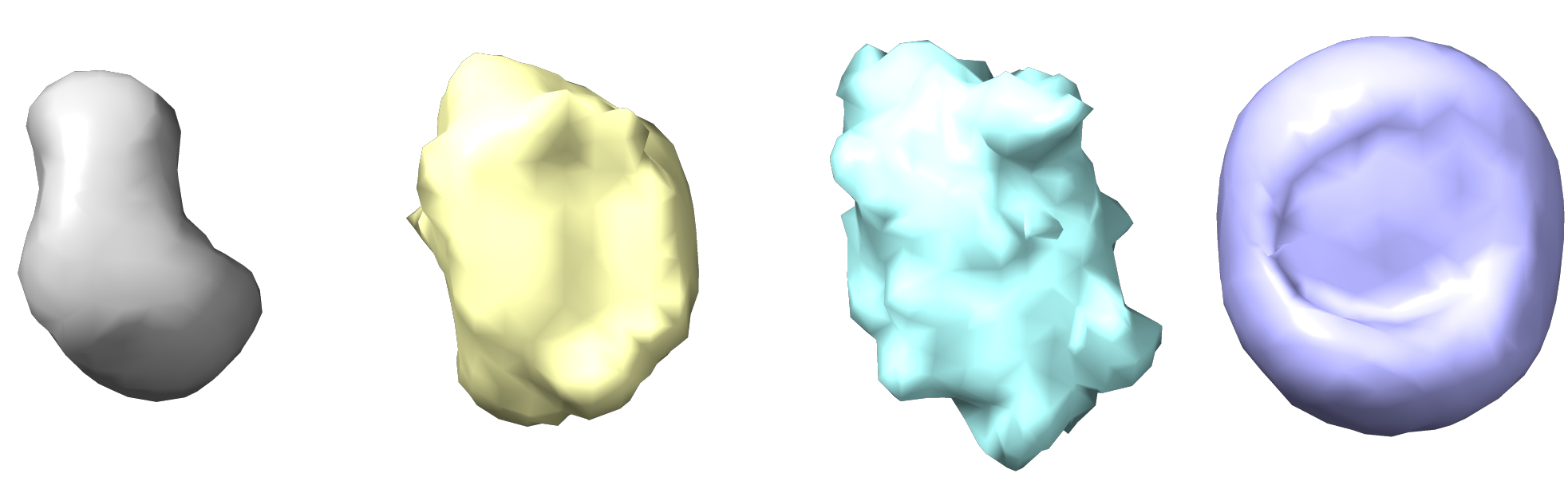}	
		\label{fig:BPTI_comp_2}} \vfill
	\subfloat[The FSC curves of the volumes in panel (A), with respect to the ground truth.]{
		\includegraphics[width=0.50\columnwidth, keepaspectratio]{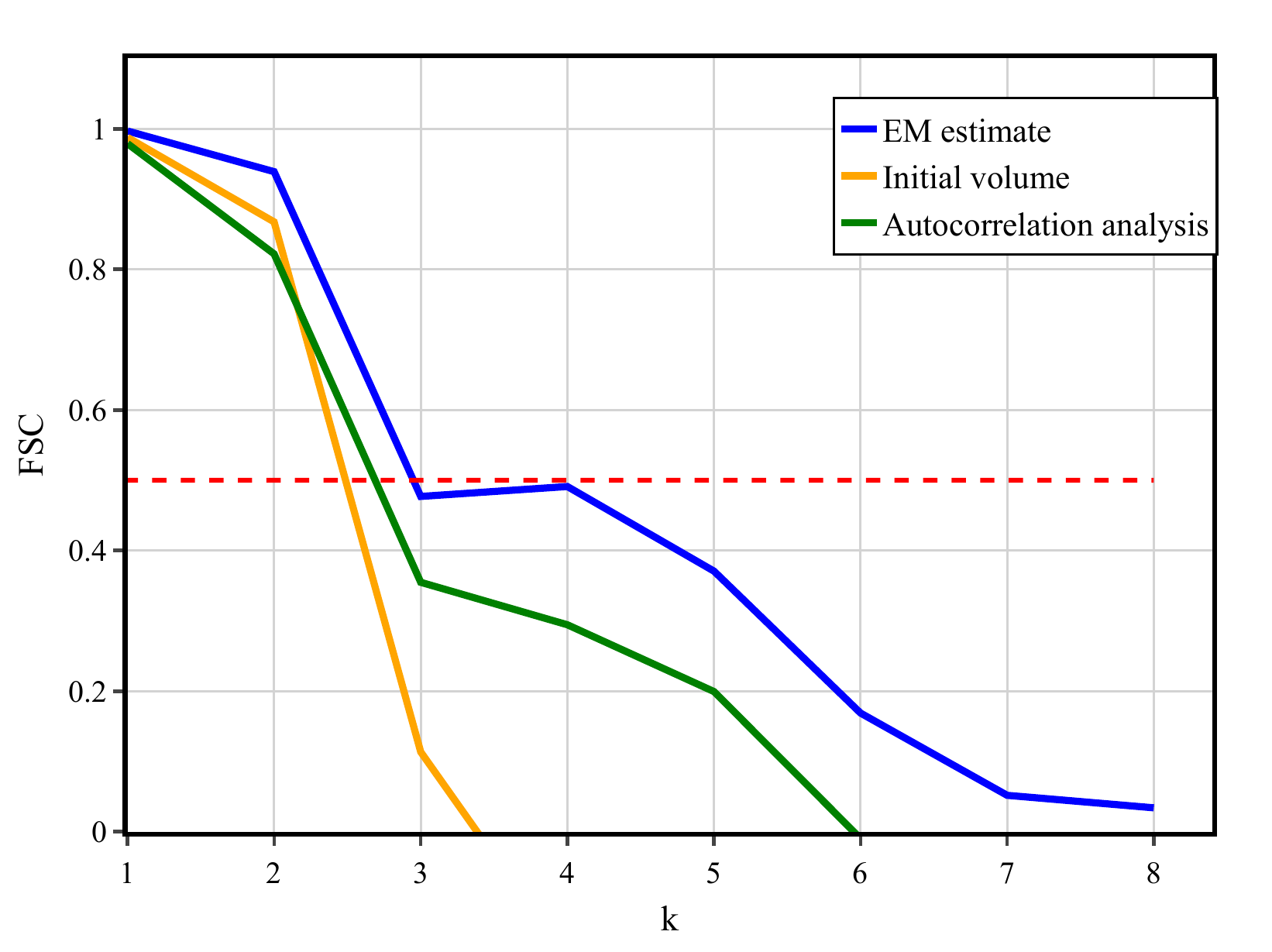}
		\label{fig:BPTI_FSC_2}}
	\caption{Results for estimating the BPTI mutant directly from a micrograph.
\label{fig:BPTI_results_2}}
\end{figure}

\subsubsection[The Bovine Pancreatic Trypsin Inhibitor (BPTI) mutant]{The Bovine Pancreatic Trypsin Inhibitor (BPTI) mutant~\cite{czapinska2000high}}
The volume was generated in~\cite{bendory2023toward} from the atomic model available as 1QLQ in the Protein Data Bank (PDB)\footnote{\url{https://www.rcsb.org/}}. We consider a version of the volume of size~$L^3 = 17^3 \text{ voxels}$. The initial guess for the algorithm was generated from the AlphaFold Protein Structure Database (AFDB), using entry~\mbox{AF-P00974-F1}, downsampled to the size of the target volume.

Using Method 2, we generated the micrograph with~$\SNR = 6.2$. A visual comparison between the estimate and the ground truth is presented in Figure~\ref{fig:BPTI_comp_2}, and the FSC curve is provided in Figure~\ref{fig:BPTI_FSC_2}. The recovery of the BPTI mutant is less successful compared to the previous volumes, but still represents a significant improvement over the initial guess.

Next, we compare our algorithm's results to those of~\cite{bendory2023toward} using autocorrelation analysis; Figure~\ref{fig:BPTI_comp_2} presents a visual comparison between our estimate and the downsampled estimate from~\cite{bendory2023toward}, and Figure~\ref{fig:BPTI_FSC_2} presents the FSC comparison between both estimates, with respect to the ground truth. The recovery in~\cite{bendory2023toward} was performed from clean autocorrelations, corresponding to~$N \rightarrow \infty$, whereas our reconstructions were obtained from finite, noisy micrographs. We note that we present the volume as originally estimated in~\cite{bendory2023toward}, initialized from a random Gaussian inital guess; initializing from the initial guess used for our algorithm did not meaningfully affect the reconstruction. That recovery was also performed for a larger volume of size~$L^3 = 31^3 \text{ voxels}$. 
It is evident that our likelihood-based method yields superior reconstruction quality. Under mild regularity conditions, the maximum likelihood estimator (MLE) is efficient---that is, it achieves the Cram\'{e}r--Rao lower bound and is asymptotically optimal; this property does not generally hold for autocorrelation analysis, which can be interpreted as a low-order approximation to it~\cite{katsevich2023likelihood}. However, our algorithm treats patches as independent, meaning the estimator does not maximize the true likelihood. Both approaches are therefore suboptimal compared to the true MLE, and it is not clear a priori which approximation is more costly. Recent results on a related model suggest the independence approximation preserves most of the statistical information~\cite{abraham2025sample}, but the comparison is ultimately an empirical one.
}

\subsection{Arbitrary spacing distribution of projection images within the micrograph}
\label{subsec:arbitrary_spacing_distribution}
\begin{figure}[t]
	\centering
	\includegraphics[width=0.35\columnwidth, keepaspectratio]{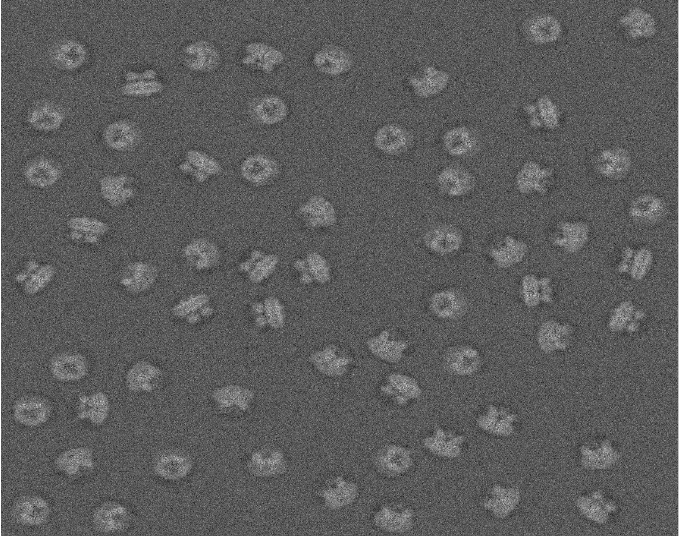}%
	\caption{A micrograph with an arbitrary spacing distribution of projection images, with~$\SNR = 1$. In contrast to Figure~\ref{fig:example_micrographs}, here the projection images can be arbitrarily close (but do not overlap).}
	\label{fig:micrograph_arbitrary_spacing_distribution}
\end{figure}

So far, we considered the case of well-separated micrographs, where at least a full projection length separates each projection image from its neighbor~\eqref{eq:sep}. This allowed us to develop the approximate EM framework (Algorithm~\ref{alg:stochastic_approximate_EM}) under the assumption that each patch contains at most one projection image. We now discuss the case of an arbitrary spacing distribution of projection images within the micrograph, which better reflects the spacing distribution in practical cryo-EM micrographs. In this model, we assume only that the projection images do not overlap; see Figure~\ref{fig:micrograph_arbitrary_spacing_distribution} for an example of such simulated micrograph. Consequently, when partioning the measurement to~$N_{\text{patches}}$ of size~$L \times L$, each patch can now consist of up to 4 projection images. In order to integrate the arbitrary spacing distribution model into the approximate EM framework, one needs to account for the possibility of multiple projections in a single patch. Namely, the set of possible shifts within a patch of all four potential image projections is~$\rev{\mathbb{S}}^4$, and the space of possible rotations is the group~\mbox{$\SO(3) \times \SO(3) \times \SO(3) \times \SO(3)$}. A similar analysis was conducted for 1-D~\cite{lan2020multi} and 2-D~\cite{kreymer2022two} related models. Clearly, this mechanism will greatly inflate the computational complexity of the algorithm.

\begin{figure}[t]
\centering
	\subfloat[Volume reconstructions: Left: the initial guess; middle: the estimate up to~$\ell = 14$; right: the ground truth volume.]{
		\includegraphics[width=0.49\columnwidth, keepaspectratio]{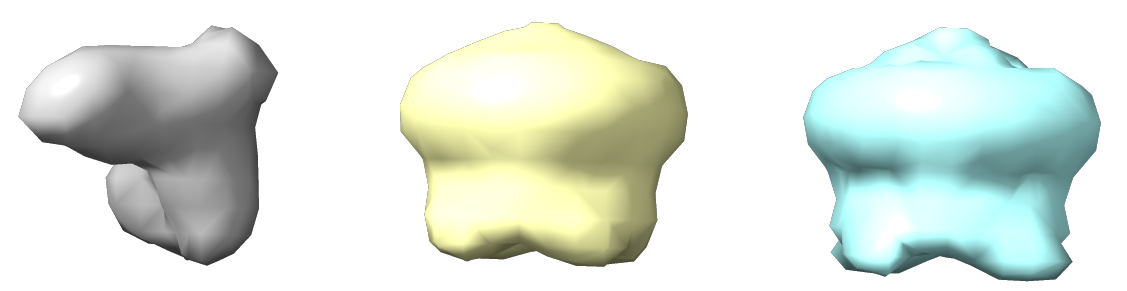}	
		\label{fig:TRPV1_comp_arbitrary_spacing_distribution}} \vfill
	\subfloat[The FSC curves of the volumes in panel (A), with respect to the ground truth.]{
		\includegraphics[width=0.50\columnwidth, keepaspectratio]{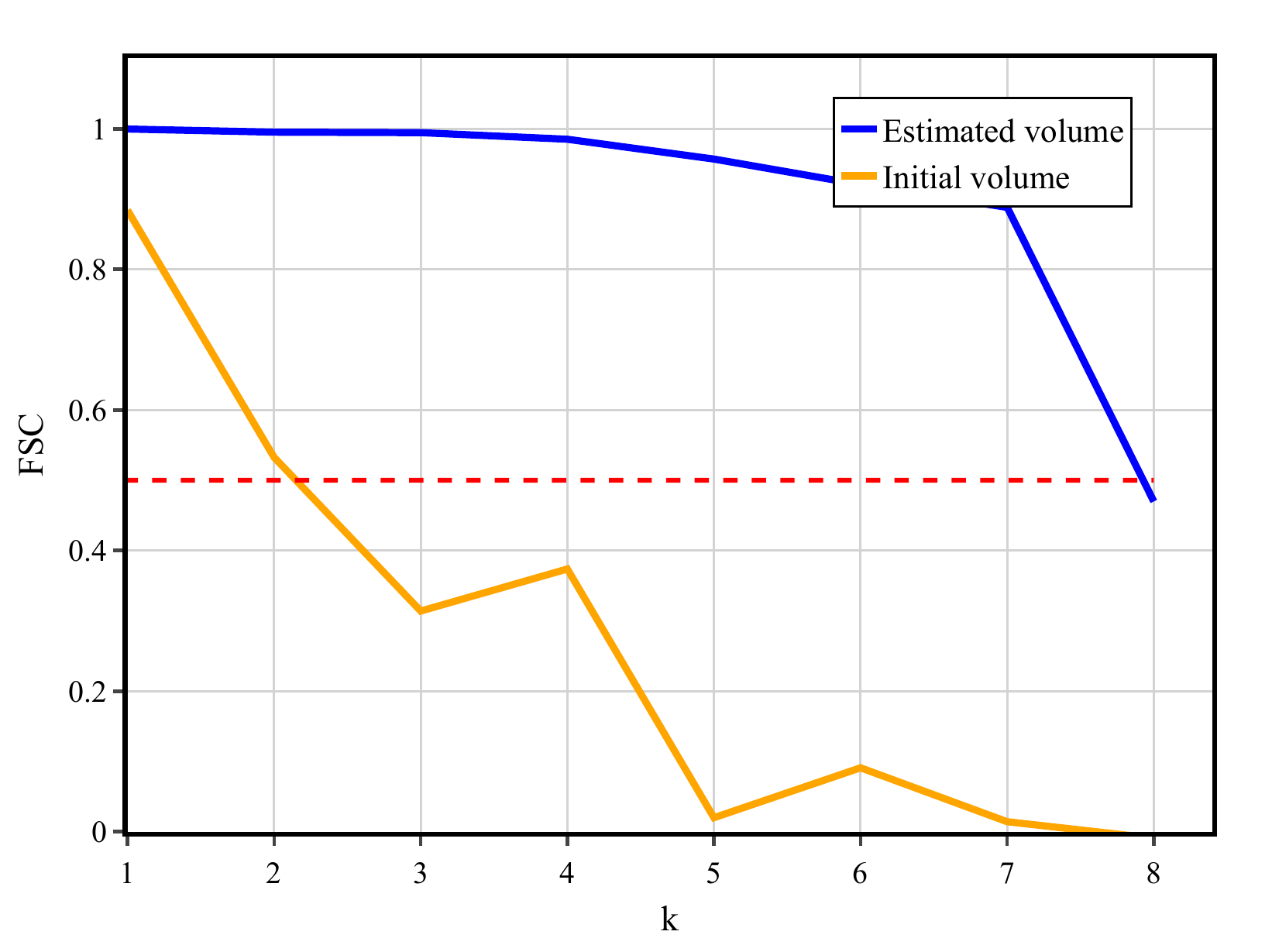}
		\label{fig:TRPV1_FSC_arbitrary_spacing_distribution}}
	\caption{Results for estimating the TRPV1 structure directly from a micrograph with an arbitrary spacing distribution of projection images.
		\label{fig:TRPV1_results_arbitrary_spacing_distribution}}
\end{figure}

In the following experiment, we examine the performance of Algorithm~\ref{alg:stochastic_approximate_EM} (that assumes well-separated projections) on micrographs with an arbitrary spacing distribution of projection images. Approximately half of the non-empty patches contain parts of more than one projection image. Nevertheless, as depicted in Figure~\ref{fig:TRPV1_results_arbitrary_spacing_distribution}, we achieve an accurate estimate of the TRPV1 structure from a downsampled micrograph with an arbitrary spacing distribution of projection images. This suggests that the well-separated model may be sufficient to achieve recoveries of reasonable resolution. A visual comparison between the true and estimated volumes is presented in Figure~\ref{fig:TRPV1_comp_arbitrary_spacing_distribution}, and the FSC curve is given in Figure~\ref{fig:TRPV1_FSC_arbitrary_spacing_distribution}.

\section{Conclusion}
\label{sec:conclusions}
In this paper, we demonstrated successful 3-D volume reconstructions directly from simulated cryo-EM micrographs, using the approximate EM algorithm. Our approach allows us to estimate the target volume directly from the measurement, without the need for particle picking. Therefore, it might be possible to reconstruct small molecular structures, and in particular structures that are too small to be detected by particle picking methods.

\paragraph{Model extensions.} Our cryo-EM micrograph generation model (see Section~\ref{sec:math_model}) is simplified and incomplete. As was mentioned in Section~\ref{subsec:arbitrary_spacing_distribution}, practical cryo-EM micrographs do not follow the well-separated model~\eqref{eq:sep}, but rather have an arbitrary spacing distribution of projection images. An initial experiment (see Figure~\ref{fig:TRPV1_results_arbitrary_spacing_distribution}) suggests that perhaps applying the stochastic approximate EM algorithm also to those micrographs will result in sufficiently accurate estimates. Moreover, the noise in experimental micrographs may be colored, and the viewing angles of the projection images are typically not uniformly distributed. \rev{Incorporating a non-uniform prior over the viewing-angle distribution is natural in our framework---the uniform prior can be replaced by a distribution over rotations, which can be estimated jointly within the EM iterations. Additionally, experimental micrographs often contain contaminants or junk particles that do not correspond to the target molecule; this is a well-known challenge in the field~\cite{agard2014single, langlois2014automated, dhakal2023large}. A possible extension is to introduce an explicit "junk class" into the generative model, which might allow the algorithm to absorb poorly explained patches rather than letting them degrade the volume estimate.} The cryo-EM measurement is also affected by the electron microscope's point spread function~\cite{erickson1971measurement}\rev{, whose effect in Fourier space is captured by the contrast transfer function (CTF). The CTF can be incorporated into our framework by modulating the projection images in Fourier space prior to the likelihood computation, as is standard in existing cryo-EM reconstruction software~\cite{scheres2012relion, punjani2017cryosparc}. Moreover, the CTF parameters can be estimated jointly with the volume within the EM iterations, by treating them as additional latent variables}. Addressing those modeling issues is essential to apply the proposed algorithm to experimental data sets.

\paragraph{Computational acceleration.} Recovering structures from highly noisy micrographs\rev{---specifically those where traditional particle-picking methods fail---}would require processing significantly more data. \rev{While the experiments in this work were conducted at noise levels where particle picking may still be viable, they serve as a proof of concept for the methodological and algorithmic novelty of reconstructing directly from micrographs.} The main hurdle to achieving this goal \rev{in higher noise regimes} is the computational burden of the algorithm. In this paper, we accelerated the algorithm by using a frequency marching scheme and a stochastic strategy. Nevertheless, further acceleration measures should be considered. A possible approach is to design a branch-and-bound algorithm that efficiently rules out regions of the search space with low probability to maximize the likelihood function~\cite{punjani2017cryosparc}. In addition, the EM iterations could be initialized by efficient computational techniques, such as autocorrelation analysis~\cite{bendory2023toward}.

\paragraph{Initialization and data-driven priors.} We used the output of AlphaFold~\cite{jumper2021highly}, a program designed using deep learning techniques to predict protein structures from their amino acid sequences, as the initial guess to our algorithm. \rev{Since the molecule's amino acid sequence is typically known, an AlphaFold prediction is readily available. Although these predictions are not always perfect,} this approach significantly improved our accuracy---our algorithm generally did not converge to a correct estimation from a random initial guess---and reduced the number of required iterations, and by that alleviated the computational complexity. In light of this result, we suggest using data-driven diffusion priors, derived from score-based stochastic differential equation models~\cite{song2022solving}, and incorporating them into the EM scheme. Adding a prior (in the Bayesian sense) to the 3-D volume is natural in the EM algorithm. \rev{Such generative priors can also serve as robust initializations, particularly in cases where AlphaFold predictions are insufficient.} In~\cite{zabatani2024score}, we successfully applied this approach to a simplified 2-D model of the problem. Extending this methodology to the 3-D setting is an ongoing research direction. \rev{Another promising direction is to replace the Fourier-Bessel expansion~\eqref{eq:3-d-fourier-bessel} with a data-driven basis obtained via $\SO(3)$-invariant PCA~\cite{fraiman2025so}. Such a basis, learned from a dataset of molecular volumes, could provide a more compact representation of the volume, potentially simplifying the EM iterations and reducing the computational complexity.}

\section*{Acknowledgments}
Shay Kreymer is supported by TAD Excellence Program for Doctoral Students in Artificial Intelligence and Data Science. Amit Singer is supported in part by AFOSR awards FA9550-20-1-0266 and FA9550-23-1-0249, in part by Simons Foundation Math+X Investigator Award, in part by NSF awards DMS-2009753 and DMS-2510039, and in part by NIH/NIGMS under Grant R01GM136780-01. Tamir Bendory is supported in part by BSF under Grant 2020159, in part by NSF-BSF under Grant 2019752, and in part by ISF under Grant 1924/21.

\rev{The authors would like to thank the anonymous reviewers for their insightful comments and constructive feedback, which have significantly improved the quality of this manuscript.} The authors also thank Eitan Levin for his helpful assistance during the initial stages of code implementation, \rev{Rafi Beinhorn for his help in reconstructing the BPTI volume using autocorrelation analysis}, and Alon Zabatani for his help in generating 3-D volumes from the GEOM dataset.

\bibliographystyle{abbrv}
\bibliography{references}

@article{lan2020multi,
  title={Multi-target detection with an arbitrary spacing distribution},
  author={Lan, Ti-Yen and Bendory, Tamir and Boumal, Nicolas and Singer, Amit},
  journal={IEEE Transactions on Signal Processing},
  volume={68},
  pages={1589--1601},
  year={2020},
  publisher={IEEE}
}

@article{bendory2020single,
  title={Single-particle cryo-electron microscopy: Mathematical theory, computational challenges, and opportunities},
  author={Bendory, Tamir and Bartesaghi, Alberto and Singer, Amit},
  journal={IEEE Signal Processing Magazine},
  volume={37},
  number={2},
  pages={58--76},
  year={2020},
  publisher={IEEE}
}

@article{bendory2023toward,
  title={Toward single particle reconstruction without particle picking: Breaking the detection limit},
  author={Bendory, Tamir and Boumal, Nicolas and Leeb, William and Levin, Eitan and Singer, Amit},
  journal={SIAM Journal on Imaging Sciences},
  volume={16},
  number={2},
  pages={886--910},
  year={2023},
  publisher={SIAM}
}

@article{bendory2019multi,
  title={Multi-target detection with application to cryo-electron microscopy},
  author={Bendory, Tamir and Boumal, Nicolas and Leeb, William and Levin, Eitan and Singer, Amit},
  journal={Inverse Problems},
  volume={35},
  number={10},
  pages={104003},
  year={2019},
  publisher={IOP Publishing}
}

@article{punjani2017cryosparc,
  title={cryo{SPARC}: Algorithms for rapid unsupervised cryo-{EM} structure determination},
  author={Punjani, Ali and Rubinstein, John L and Fleet, David J and Brubaker, Marcus A},
  journal={Nature Methods},
  volume={14},
  number={3},
  pages={290--296},
  year={2017},
  publisher={Nature Publishing Group}
}

@inproceedings{marshall2020image,
  title={Image recovery from rotational and translational invariants},
  author={Marshall, Nicholas F and Lan, Ti-Yen and Bendory, Tamir and Singer, Amit},
  booktitle={ICASSP 2020-2020 IEEE International Conference on Acoustics, Speech and Signal Processing (ICASSP)},
  pages={5780--5784},
  year={2020},
  organization={IEEE}
}

@article{bai2015cryo,
  title={How cryo-{EM} is revolutionizing structural biology},
  author={Bai, Xiao-Chen and McMullan, Greg and Scheres, Sjors HW},
  journal={Trends in Biochemical Sciences},
  volume={40},
  number={1},
  pages={49--57},
  year={2015},
  publisher={Elsevier}
}

@article{sigworth1998maximum,
  title={A maximum-likelihood approach to single-particle image refinement},
  author={Sigworth, Fred J},
  journal={Journal of structural biology},
  volume={122},
  number={3},
  pages={328--339},
  year={1998},
  publisher={Elsevier}
}

@article{nogales2016development,
  title={The development of cryo-{EM} into a mainstream structural biology technique},
  author={Nogales, Eva},
  journal={Nature Methods},
  volume={13},
  number={1},
  pages={24--27},
  year={2016},
  publisher={Nature Publishing Group}
}

@article{henderson1995potential,
  title={The potential and limitations of neutrons, electrons and {X}-rays for atomic resolution microscopy of unstained biological molecules},
  author={Henderson, Richard},
  journal={Quarterly Reviews of Biophysics},
  volume={28},
  number={2},
  pages={171--193},
  year={1995},
  publisher={Cambridge University Press}
}

@book{frank2006three,
  title={Three-dimensional electron microscopy of macromolecular assemblies: Visualization of biological molecules in their native state},
  author={Frank, Joachim},
  year={2006},
  publisher={Oxford University Press}
}

@article{aguerrebere2016fundamental,
  title={Fundamental limits in multi-image alignment},
  author={Aguerrebere, Cecilia and Delbracio, Mauricio and Bartesaghi, Alberto and Sapiro, Guillermo},
  journal={IEEE Transactions on Signal Processing},
  volume={64},
  number={21},
  pages={5707--5722},
  year={2016},
  publisher={IEEE}
}

@article{bendory2023multi,
title = {Multi-target detection with rotations},
journal = {Inverse Problems and Imaging},
volume = {17},
number = {2},
pages = {362-380},
year = {2023},
author = {Tamir Bendory and Ti-Yen Lan and Nicholas F. Marshall and Iris Rukshin and Amit Singer},
}

@article{scheres2012relion,
  title={{RELION}: implementation of a {B}ayesian approach to cryo-{EM} structure determination},
  author={Scheres, Sjors HW},
  journal={Journal of Structural Biology},
  volume={180},
  number={3},
  pages={519--530},
  year={2012},
  publisher={Elsevier}
}

@article{kreymer2022two,
  title={Two-dimensional multi-target detection: An autocorrelation analysis approach},
  author={Kreymer, Shay and Bendory, Tamir},
  journal={IEEE Transactions on Signal Processing},
  volume={70},
  pages={835--849},
  year={2022},
  publisher={IEEE}
}

@article{dempster1977maximum,
  title={Maximum likelihood from incomplete data via the {EM} algorithm},
  author={Dempster, Arthur P and Laird, Nan M and Rubin, Donald B},
  journal={Journal of the Royal Statistical Society: Series B (Methodological)},
  volume={39},
  number={1},
  pages={1--22},
  year={1977},
  publisher={Wiley Online Library}
}

@article{katsevich2023likelihood,
  title={Likelihood maximization and moment matching in low {SNR} {G}aussian mixture models},
  author={Katsevich, Anya and Bandeira, Afonso S},
  journal={Communications on Pure and Applied Mathematics},
  volume={76},
  number={4},
  pages={788--842},
  year={2023},
  publisher={Wiley Online Library}
}

@inproceedings{shalit2022generalized,
  title={Generalized Autocorrelation Analysis for Multi-Target Detection},
  author={Shalit, Ye'Ela and Weber, Ran and Abas, Asaf and Kreymer, Shay and Bendory, Tamir},
  booktitle={ICASSP 2022-2022 IEEE International Conference on Acoustics, Speech and Signal Processing (ICASSP)},
  pages={5907--5911},
  year={2022},
  organization={IEEE}
}

@article{singer2020computational,
  title={Computational methods for single-particle electron cryomicroscopy},
  author={Singer, Amit and Sigworth, Fred J},
  journal={Annual Review of Biomedical Data Science},
  volume={3},
  pages={163--190},
  year={2020},
  publisher={Annual Reviews}
}

@article{kreymer2022approximate,
  title={An approximate expectation-maximization for two-dimensional multi-target detection},
  author={Kreymer, Shay and Singer, Amit and Bendory, Tamir},
  journal={IEEE Signal Processing Letters},
  volume={29},
  pages={1087--1091},
  year={2022},
  publisher={IEEE}
}

@inproceedings{levin20183d,
  title={{3D} ab initio modeling in cryo-{EM} by autocorrelation analysis},
  author={Levin, Eitan and Bendory, Tamir and Boumal, Nicolas and Kileel, Joe and Singer, Amit},
  booktitle={2018 IEEE 15th International Symposium on Biomedical Imaging (ISBI 2018)},
  pages={1569--1573},
  year={2018},
  organization={IEEE}
}

@article{bhamre2017anisotropic,
  title={Anisotropic twicing for single particle reconstruction using autocorrelation analysis},
  author={Bhamre, Tejal and Zhang, Teng and Singer, Amit},
  journal={arXiv preprint arXiv:1704.07969},
  year={2017}
}

@book{natterer2001mathematics,
author = {Natterer, Frank},
title = {The Mathematics of Computerized Tomography},
year = {2001},
isbn = {0898714931},
publisher = {Society for Industrial and Applied Mathematics},
address = {USA}
}

@article{slepian1964prolate,
  title={Prolate spheroidal wave functions, {F}ourier analysis and uncertainty—{IV}: Extensions to many dimensions; generalized prolate spheroidal functions},
  author={Slepian, David},
  journal={Bell System Technical Journal},
  volume={43},
  number={6},
  pages={3009--3057},
  year={1964},
  publisher={Wiley Online Library}
}

@article{landa2017steerable,
  title={Steerable principal components for space-frequency localized images},
  author={Landa, Boris and Shkolnisky, Yoel},
  journal={SIAM Journal on Imaging Sciences},
  volume={10},
  number={2},
  pages={508--534},
  year={2017},
  publisher={SIAM}
}

@article{eldar2020klt,
  title={{KLT} picker: Particle picking using data-driven optimal templates},
  author={Eldar, Amitay and Landa, Boris and Shkolnisky, Yoel},
  journal={Journal of Structural Biology},
  volume={210},
  number={2},
  pages={107473},
  year={2020},
  publisher={Elsevier}
}

@article{heimowitz2018apple,
  title={{APPLE} picker: Automatic particle picking, a low-effort cryo-{EM} framework},
  author={Heimowitz, Ayelet and And{\'e}n, Joakim and Singer, Amit},
  journal={Journal of Structural Biology},
  volume={204},
  number={2},
  pages={215--227},
  year={2018},
  publisher={Elsevier}
}

@article{bepler2019positive,
  title={Positive-unlabeled convolutional neural networks for particle picking in cryo-electron micrographs},
  author={Bepler, Tristan and Morin, Andrew and Rapp, Micah and Brasch, Julia and Shapiro, Lawrence and Noble, Alex J and Berger, Bonnie},
  journal={Nature Methods},
  volume={16},
  number={11},
  pages={1153--1160},
  year={2019},
  publisher={Nature Publishing Group}
}

@article{wang2016deeppicker,
  title={Deep{P}icker: A deep learning approach for fully automated particle picking in cryo-{EM}},
  author={Wang, Feng and Gong, Huichao and Liu, Gaochao and Li, Meijing and Yan, Chuangye and Xia, Tian and Li, Xueming and Zeng, Jianyang},
  journal={Journal of Structural Biology},
  volume={195},
  number={3},
  pages={325--336},
  year={2016},
  publisher={Elsevier}
}

@article{sigworth2004classical,
  title={Classical detection theory and the cryo-{EM} particle selection problem},
  author={Sigworth, Fred J},
  journal={Journal of Structural Biology},
  volume={145},
  number={1-2},
  pages={111--122},
  year={2004},
  publisher={Elsevier}
}

@article{glaeser1999electron,
  title={Electron crystallography: Present excitement, a nod to the past, anticipating the future},
  author={Glaeser, Robert M},
  journal={Journal of Structural Biology},
  volume={128},
  number={1},
  pages={3--14},
  year={1999},
  publisher={Elsevier}
}

@article{scapin2018cryo,
  title={Cryo-{EM} for small molecules discovery, design, understanding, and application},
  author={Scapin, Giovanna and Potter, Clinton S and Carragher, Bridget},
  journal={Cell Chemical Biology},
  volume={25},
  number={11},
  pages={1318--1325},
  year={2018},
  publisher={Elsevier}
}

@article{zhang2019cryo,
  title={Cryo-{EM} structure of a 40 k{D}a {SAM}-{IV} riboswitch {RNA} at 3.7 {{\AA}} resolution},
  author={Zhang, Kaiming and Li, Shanshan and Kappel, Kalli and Pintilie, Grigore and Su, Zhaoming and Mou, Tung-Chung and Schmid, Michael F and Das, Rhiju and Chiu, Wah},
  journal={Nature Communications},
  volume={10},
  number={1},
  pages={1--6},
  year={2019},
  publisher={Nature Publishing Group}
}

@article{wu2020low,
  title={How low can we go? {S}tructure determination of small biological complexes using single-particle cryo-{EM}},
  author={Wu, Mengyu and Lander, Gabriel C},
  journal={Current Opinion in Structural Biology},
  volume={64},
  pages={9--16},
  year={2020},
  publisher={Elsevier}
}

@article{bai2021seeing,
  title={Seeing atoms by single-particle cryo-{EM}},
  author={Bai, Xiao-Chen},
  journal={Trends in Biochemical Sciences},
  volume={46},
  number={4},
  pages={253--254},
  year={2021},
  publisher={Elsevier}
}

@article{liu20193,
  title={A 3.8 {{\AA}} resolution cryo-{EM} structure of a small protein bound to an imaging scaffold},
  author={Liu, Yuxi and Huynh, Duc T and Yeates, Todd O},
  journal={Nature Communications},
  volume={10},
  number={1},
  pages={1--7},
  year={2019},
  publisher={Nature Publishing Group}
}

@article{yeates2020development,
  title={Development of imaging scaffolds for cryo-electron microscopy},
  author={Yeates, Todd O and Agdanowski, Matthew P and Liu, Yuxi},
  journal={Current Opinion in Structural Biology},
  volume={60},
  pages={142--149},
  year={2020},
  publisher={Elsevier}
}

@article{wu2012fabs,
  title={Fabs enable single particle cryo{EM} studies of small proteins},
  author={Shenping, Wu and Agustin, Avila-Sakar and JungMin, Kim and David S., Booth and Charles H., Greenberg and Andrea, Rossi and others},
  journal={Structure},
  volume={20},
  number={4},
  pages={582--592},
  year={2012},
  publisher={Elsevier}
}

@article{wu2021cryo,
  title={Cryo-{EM} structure determination of small proteins by nanobody-binding scaffolds ({L}egobodies)},
  author={Wu, Xudong and Rapoport, Tom A},
  journal={Proceedings of the National Academy of Sciences},
  volume={118},
  number={41},
  pages={e2115001118},
  year={2021},
  publisher={National Acad Sciences}
}

@article{danev2017expanding,
  title={Expanding the boundaries of cryo-{EM} with phase plates},
  author={Danev, Radostin and Baumeister, Wolfgang},
  journal={Current Opinion in Structural Biology},
  volume={46},
  pages={87--94},
  year={2017},
  publisher={Elsevier}
}

@article{freeman1991design,
  title={The design and use of steerable filters},
  author={Freeman, William T and Adelson, Edward H},
  journal={IEEE Transactions on Pattern Analysis and Machine Intelligence},
  volume={13},
  number={9},
  pages={891--906},
  year={1991}
}

@article{hinton2006fast,
  title={A fast learning algorithm for deep belief nets},
  author={Hinton, Geoffrey E and Osindero, Simon and Teh, Yee-Whye},
  journal={Neural Computation},
  volume={18},
  number={7},
  pages={1527--1554},
  year={2006},
  publisher={MIT Press One Rogers Street, Cambridge, MA 02142-1209, USA journals-info~…}
}

@article{hyvarinen2013independent,
  title={Independent component analysis: Recent advances},
  author={Hyv{\"a}rinen, Aapo},
  journal={Philosophical Transactions of the Royal Society A: Mathematical, Physical and Engineering Sciences},
  volume={371},
  number={1984},
  pages={20110534},
  year={2013},
  publisher={The Royal Society Publishing}
}

@article{segol2021improved,
  title={Improved convergence guarantees for learning {G}aussian mixture models by {EM} and gradient {EM}},
  author={Segol, Nimrod and Nadler, Boaz},
  journal={Electronic Journal of Statistics},
  volume={15},
  number={2},
  pages={4510--4544},
  year={2021},
  publisher={Institute of Mathematical Statistics and Bernoulli Society}
}

@incollection{neal1998view,
  title={A view of the {EM} algorithm that justifies incremental, sparse, and other variants},
  author={Neal, Radford M and Hinton, Geoffrey E},
  booktitle={Learning in Graphical Models},
  pages={355--368},
  year={1998},
  publisher={Springer}
}

@article{barnett2017rapid,
  title={Rapid solution of the cryo-{EM} reconstruction problem by frequency marching},
  author={Barnett, Alex and Greengard, Leslie and Pataki, Andras and Spivak, Marina},
  journal={SIAM Journal on Imaging Sciences},
  volume={10},
  number={3},
  pages={1170--1195},
  year={2017},
  publisher={SIAM}
}

@article{gao2016trpv1,
  title={{TRPV1} structures in nanodiscs reveal mechanisms of ligand and lipid action},
  author={Gao, Yuan and Cao, Erhu and Julius, David and Cheng, Yifan},
  journal={Nature},
  volume={534},
  number={7607},
  pages={347--351},
  year={2016},
  publisher={Nature Publishing Group}
}

@article{czapinska2000high,
  title={High-resolution structure of bovine pancreatic trypsin inhibitor with altered binding loop sequence},
  author={Czapinska, Honorata and Otlewski, Jacek and Krzywda, Szymon and Sheldrick, George M and Jask{\'o}lski, Mariusz},
  journal={Journal of Molecular Biology},
  volume={295},
  number={5},
  pages={1237--1249},
  year={2000},
  publisher={Elsevier}
}

@article{sigworth2016principles,
  title={Principles of cryo-{EM} single-particle image processing},
  author={Sigworth, Fred J},
  journal={Microscopy},
  volume={65},
  number={1},
  pages={57--67},
  year={2016},
  publisher={Oxford University Press}
}

@incollection{shoemake1992uniform,
  title={Uniform random rotations},
  author={Shoemake, Ken},
  booktitle={Graphics Gems III (IBM Version)},
  pages={124--132},
  year={1992},
  publisher={Elsevier}
}

@article{pettersen2004ucsf,
  title={{UCSF} {C}himera—a visualization system for exploratory research and analysis},
  author={Pettersen, Eric F and Goddard, Thomas D and Huang, Conrad C and Couch, Gregory S and Greenblatt, Daniel M and Meng, Elaine C and Ferrin, Thomas E},
  journal={Journal of Computational Chemistry},
  volume={25},
  number={13},
  pages={1605--1612},
  year={2004},
  publisher={Wiley Online Library}
}

@article{elmlund2015cryogenic,
  title={Cryogenic electron microscopy and single-particle analysis},
  author={Elmlund, Dominika and Elmlund, Hans},
  journal={Annual Review of Biochemistry},
  volume={84},
  number={1},
  pages={499--517},
  year={2015}
}

@article{cheng2015primer,
  title={A primer to single-particle cryo-electron microscopy},
  author={Cheng, Yifan and Grigorieff, Nikolaus and Penczek, Pawel A and Walz, Thomas},
  journal={Cell},
  volume={161},
  number={3},
  pages={438--449},
  year={2015},
  publisher={Elsevier}
}

@article{erickson1971measurement,
  title={Measurement and compensation of defocusing and aberrations by {F}ourier processing of electron micrographs},
  author={Erickson, HP and Klug, Aaron},
  journal={Philosophical Transactions of the Royal Society of London. B, Biological Sciences},
  volume={261},
  number={837},
  pages={105--118},
  year={1971},
  publisher={The Royal Society London}
}

@article{zheng2022uniform,
  title={Uniform thin ice on ultraflat graphene for high-resolution cryo-{EM}},
  author={Zheng, Liming and
Liu, Nan and
Gao, Xiaoyin and
Zhu, Wenqing and
Liu, Kun and
Wu, Cang and others},
  journal={Nature Methods},
  pages={1--8},
  year={2022},
  publisher={Nature Publishing Group}
}

@article{feder1988parameter,
	title={Parameter estimation of superimposed signals using the {EM} algorithm},
	author={Feder, Meir and Weinstein, Ehud},
	journal={IEEE Transactions on Acoustics, Speech, and Signal Processing},
	volume={36},
	number={4},
	pages={477--489},
	year={1988},
	publisher={IEEE}
}

@article{neyman1948consistent,
	title={Consistent estimates based on partially consistent observations},
	author={Neyman, Jerzy and Scott, Elizabeth L},
	journal={Econometrica: Journal of the Econometric Society},
	pages={1--32},
	year={1948},
	publisher={JSTOR}
}

@inproceedings{song2022solving,
  author    = {Yang Song and
               Liyue Shen and
               Lei Xing and
               Stefano Ermon},
  title     = {Solving Inverse Problems in Medical Imaging with Score-Based Generative
               Models},
  booktitle = {The Tenth International Conference on Learning Representations, {ICLR} 2022},
  year      = {2022}
}

@misc{garrett_wright_2023_7510635,
  author       = {Garrett Wright and
                  Joakim Andén and
                  Vineet Bansal and
                  Junchao Xia and
                  Chris Langfield and
                  Josh Carmichael and
                  Robbie Brook and
                  Yunpeng Shi and
                  Ayelet Heimowitz and
                  Gabi Pragier and
                  Itay Sason and
                  Amit Moscovich and
                  Yoel Shkolnisky and
                  Amit Singer},
  title        = {Computational{C}ryo{EM}/{ASPIRE}-{P}ython: v0.10.1},
  month        = jan,
  year         = 2023,
  publisher    = {Zenodo},
  version      = {v0.10.1},
  doi          = {10.5281/zenodo.5657281},
}

@article{harauz1983direct,
  title={Direct three-dimensional reconstruction for macromolecular complexes from electron micrographs},
  author={Harauz, George and Ottensmeyer, FP},
  journal={Ultramicroscopy},
  volume={12},
  number={4},
  pages={309--319},
  year={1983},
  publisher={Elsevier}
}

@article{jumper2021highly,
	title={Highly accurate protein structure prediction with {A}lpha{F}old},
	author={Jumper, John and Evans, Richard and Pritzel, Alexander and Green, Tim and Figurnov, Michael and others},
	journal={Nature},
	volume={596},
	number={7873},
	pages={583--589},
	year={2021},
	publisher={Nature Publishing Group}
}

@article{varadi2024alphafold,
	title={{A}lpha{F}old {P}rotein {S}tructure {D}atabase in 2024: providing structure coverage for over 214 million protein sequences},
	author={Varadi, Mihaly and Bertoni, Damian and Magana, Paulyna and Paramval, Urmila and Pidruchna, Ivanna and others},
	journal={Nucleic Acids Research},
	volume={52},
	number={D1},
	pages={D368--D375},
	year={2024},
	publisher={Oxford University Press}
}

@article{varadi2022alphafold,
	title={{A}lpha{F}old {P}rotein {S}tructure {D}atabase: massively expanding the structural coverage of protein-sequence space with high-accuracy models},
	author={Varadi, Mihaly and Anyango, Stephen and Deshpande, Mandar and Nair, Sreenath and Natassia, Cindy and others},
	journal={Nucleic Acids Research},
	volume={50},
	number={D1},
	pages={D439--D444},
	year={2022},
	publisher={Oxford University Press}
}

@inproceedings{zabatani2024score,
	title={Score-based diffusion priors for multi-target detection},
	author={Zabatani, Alon and Kreymer, Shay and Bendory, Tamir},
	booktitle={2024 58th Annual Conference on Information Sciences and Systems (CISS)},
	pages={1--6},
	year={2024},
	organization={IEEE}
}

@article{axelrod2022geom,
  title={{GEOM}, energy-annotated molecular conformations for property prediction and molecular generation},
  author={Axelrod, Simon and Gomez-Bombarelli, Rafael},
  journal={Scientific Data},
  volume={9},
  number={1},
  pages={185},
  year={2022},
  publisher={Nature Publishing Group UK London}
}

@article{kimanius2024data,
  title={Data-driven regularization lowers the size barrier of cryo-{EM} structure determination},
  author={Kimanius, Dari and Jamali, Kiarash and Wilkinson, Max E and L{\"o}vestam, Sofia and Velazhahan, Vaithish and Nakane, Takanori and Scheres, Sjors HW},
  journal={Nature Methods},
  volume={21},
  number={7},
  pages={1216--1221},
  year={2024},
  publisher={Nature Publishing Group US New York}
}

@article{harrison2023review,
  title={A review of the approaches used to solve sub-100 k{D}a membrane proteins by cryo-electron microscopy},
  author={Harrison, Peter J and Vecerkova, Tereza and Clare, Daniel K and Quigley, Andrew},
  journal={Journal of Structural Biology},
  volume={215},
  number={2},
  pages={107959},
  year={2023},
  publisher={Elsevier}
}

@article{schwartz2019laser,
  title={Laser phase plate for transmission electron microscopy},
  author={Schwartz, Osip and Axelrod, Jeremy J and Campbell, Sara L and Turnbaugh, Carter and Glaeser, Robert M and M{\"u}ller, Holger},
  journal={Nature Methods},
  volume={16},
  number={10},
  pages={1016--1020},
  year={2019},
  publisher={Nature Publishing Group US New York}
}

@article{fraiman2025so,
  title={{SO}(3)-invariant {PCA} with application to molecular data},
  author={Fraiman, Michael and Hoyos, Paulina and Bendory, Tamir and Kileel, Joe and Mickelin, Oscar and Sharon, Nir and Singer, Amit},
  journal={arXiv preprint arXiv:2510.18827},
  year={2025}
}

@article{agard2014single,
  title={Single-particle cryo-electron microscopy (cryo-{EM}): Progress, challenges, and perspectives for further improvement},
  author={Agard, David and Cheng, Yifan and Glaeser, Robert M and Subramaniam, Sriram},
  journal={Advances in Imaging and Electron Physics},
  volume={185},
  pages={113--137},
  year={2014},
  publisher={Elsevier}
}

@article{langlois2014automated,
  title={Automated particle picking for low-contrast macromolecules in cryo-electron microscopy},
  author={Langlois, Robert and Pallesen, Jesper and Ash, Jordan T and Ho, Danny Nam and Rubinstein, John L and Frank, Joachim},
  journal={Journal of Structural Biology},
  volume={186},
  number={1},
  pages={1--7},
  year={2014},
  publisher={Elsevier}
}

@article{dhakal2023large,
  title={A large expert-curated cryo-{EM} image dataset for machine learning protein particle picking},
  author={Dhakal, Ashwin and Gyawali, Rajan and Wang, Liguo and Cheng, Jianlin},
  journal={Scientific Data},
  volume={10},
  number={1},
  pages={392},
  year={2023},
  publisher={Nature Publishing Group UK London}
}

@article{abraham2025sample,
  title={Sample Complexity Analysis of Multi-Target Detection via {M}arkovian and Hard-Core Multi-Reference Alignment},
  author={Abraham, Kweku and Balanov, Amnon and Bendory, Tamir and Esteve-Yag{\"u}e, Carlos},
  journal={arXiv preprint arXiv:2510.17775},
  year={2025}
}

\end{document}